\documentclass{article}
\usepackage{arxiv}
\usepackage{graphicx,amsmath,amsfonts,amscd,amssymb,pstricks}
\usepackage{dcolumn}
\usepackage{bm}
\usepackage{longtable}
\title{A novel non-adiabatic approach to transition crossing in a circular hadron accelerator}
\author{M. Giovannozzi\thanks{Corresponding author: massimo.giovannozzi@cern.ch} \\
Beams Department, CERN, Esplanade des Particules 1, 1211 Meyrin, Switzerland \\
\And
L. Huang \\
Chinese Academy of Sciences, 19B Yuquan Road, Shijingshan District, Beijing, China \\
\And
A.~Huschauer \\
Beams Department, CERN, Esplanade des Particules 1, 1211 Meyrin, Switzerland \\
\And
A. Franchi \\
ESRF, 71 Avenue des Martyrs, 38000 Grenoble, France}
\begin{document}
\maketitle
\begin{abstract}  
Crossing the transition energy is always a delicate process, representing a potential source of strong perturbations of the dynamics of charged particle beams in a hadron circular accelerator. Since the first generation of multi-GeV rings, intense studies have been devoted to understanding the possible harmful mechanisms involved in transition crossing and to devise mitigation measures. Nowadays, several circular particle accelerators are successfully operating across transition and this process is well mastered. In a completely different context, stable resonances of the traverse phase space have been proposed as new means of manipulating charged particle beams. While the original aim of such a proposal was multi-turn extraction from the CERN Proton Synchrotron to the Super Proton Synchrotron, many more applications have been proposed and studied in detail. In this paper, the two topics, i.e. transition crossing and stable resonances, have been brought together with the goal of providing a novel and non-adiabatic approach to perform a clean transition crossing. The idea presented here is that by judiciously using sextupoles and octupoles it is possible to generate stable islands of the horizontal phase space. These islands represent a second closed orbit whose properties can be selected independently of those of the standard, i.e. central, closed orbit. This provides a means of performing a non-adiabatic change of the transition energy experienced by the charged particles by displacing the beam between the two closed orbits. 
\end{abstract}
\section{INTRODUCTION}
Transition crossing is certainly a critical stage during beam acceleration in a circular particle ring. It was a great challenge for the experts at work on the beam commissioning of the CERN Proton Synchrotron (PS) ring in November 1959, till the correct way of solving the problem was found~\cite{courrier}. Crossing transition energy involves a strong perturbation of the longitudinal beam dynamics that can be enhanced by space charge and chromatic effects. All this requires several non-adiabatic measures to be applied, such as a change of RF phase and the sign of chromaticity, and all this with high-accuracy timing. 

The coupling between the transverse and longitudinal beam dynamics can be used in this context, and indeed a change of optics can be envisaged as a means to overcome some of the problems involved in transition crossing. Two techniques are recalled here, namely, the gamma-jump scheme (see, e.g. Refs.~\cite{gj1,gj2,gj3}) and the imaginary gamma-transition lattices~\cite{img1,img2}, which are lattices designed to avoid transition crossing. The former is being operationally applied at the PS for example, while the lattice of the CERN Low Energy Antiproton Ring (LEAR) was an example of the latter technique~\cite{lear}. Nowadays, several machines worldwide operate successfully through transition, among them the BNL Alternating Gradient Synchrotron (AGS) and the Relativistic Heavy Ion Collider (RHIC)~\cite{rhic1,rhic2,rhic3,rhic4}, and the CERN Super Proton Synchrotron (SPS). Attempts to manipulate the value of the transition energy dynamically to prevent crossing it in case of beam deceleration have been also reported in the literature~\cite{stancari}. This brief historical overview might give the impression that all issues related to transition crossing have been successfully overcome, and that no new developments are needed in this domain. In the rest of the paper, it will be shown that it is indeed possible to further improve the classical gamma-jump schemes with new concepts inherited from nonlinear beam dynamics. 

In recent years, a novel beam manipulation technique has been put into operation, based on beam splitting by resonance crossing in the horizontal plane~\cite{prl,epac}, and known as Multi-Turn Extraction (MTE). The process is based on the use of nonlinear beam dynamics. Stable islands are created by means of sextupole and octupole magnets, which are responsible for introducing controlled nonlinear effects in the beam dynamics. Then, an adiabatic tune variation is applied, such that a given stable resonance is crossed. During the resonance-crossing stage, particles can be trapped into the islands, and eventually transported to large amplitudes. The overall result consists of splitting the beam in the horizontal phase space, so that from the initial single Gaussian, multiple quasi-Gaussian distributions are generated (see, e.g.~\cite{prl,epac}). 

The difference between stable and unstable resonances can be summarised as follows: for an unstable resonance of order $n$ the beam is split in $n$ Gaussian beamlets with the centre of phase space almost completely depleted. On the other hand, in the case of a stable resonance of order $n$, $n+1$ Gaussian beamlets are created, including a central beamlet. It is also worth stressing the intrinsically different properties of the beamlets at nonzero amplitude with respect to the one at the origin. Indeed, while the beamlet around the origin represents a periodic structure with a periodicity of one machine turn, the other beamlets represent a single structure that winds up around the ring and closes up in a periodic way after $n$ machine turns. The central beamlet does not need to have the same properties as the external ones. This implies that an additional degree of freedom is available for the case of stable resonances when defining the protocol of resonance crossing. One can control the sharing of both emittance and intensity between the two phase-space structures. 

The technique of beam splitting was originally proposed to perform multi-turn extraction from the PS to the SPS ~\cite{prl,epac,MTE2,MTE2a,MTEDR}, but soon afterwards it was realised that many more applications could be based on resonance crossing. Indeed, this technique could be time-reversed to envisage a multiturn injection based on the merging of beamlets~\cite{MTI-PRSTAB}. Such an approach would be very appealing as it allows beam shaping, which is an interesting aspect in view of mitigating space charge effects~\cite{MTI-ipac}. Furthermore, the stability of the fourth-order resonance has been studied in detail with a method to turn it into an unstable resonance in view of generating a split beam with only four beamlets~\cite{diego}. 

Detailed experimental studies have been performed at the PS~\cite{MTE-prog} in view of an operational implementation of this novel technique to transfer the beam from the PS to the SPS~\cite{MTE-comm}. This was successfully achieved in the second half of 2015~\cite{MTEEPL}. Intense efforts were also devoted to the more theoretical aspects with the goal of understanding the details of the splitting process in a quantitative way with the help of adiabatic theory~\cite{adiab}. 

The key point that is going to be detailed is that it is possible to exploit the stable islands as an additional stable closed orbit, which can be used by the circulating beam. Moreover, it is possible to devise a means of controlling the optical properties of the stable islands such that the transition crossing can be performed by displacing the beam from the central closed orbit to the stable islands, and vice versa, which is equivalent to a sophisticated gamma-jump scheme.

Section~\ref{sec:transition}, will review the fundamental concepts related to transition-crossing manipulations, while the main concepts underlying the use of stable islands are recalled in Section~\ref{sec:splitting} with particular emphasis on the aspects that can be used for performing the novel transition-crossing scheme. In Section~\ref{sec:new_scheme} the proposed scheme is presented and discussed in detail, including the results of numerical simulations, followed by a discussion on some specific points in Section~\ref{sec:discussion}, and a conclusion. The interested reader can also find some details of the PS ring, whose lattice has been used as an example for the proposed method in the appendix.
\section{Transition crossing}\label{sec:transition}
An excellent review of the phenomena involved in the transition-crossing process can be found in Ref.~\cite{PS50V1}, with particular emphasis on the aspects relevant for the PS ring. Additional material can be found in Refs.~\cite{chao,sylee,bryant,ng}. Here, we will recall a few of the aspects of relevance.

Two aspects have to be taken into account when analysing a particle's motion during acceleration, namely the increase of particle speed and the increase of the path length along the orbit. The change of these two quantities varies in sign according to whether the motion occurs below or above a particular energy value that is called the transition energy. If one defines 
\begin{equation}
\frac{\Delta C}{C} = \alpha_{\rm c} \, \frac{\Delta p}{p},
\end{equation}
where $C$ represents the length of the particle's orbit, then $\alpha_{\rm c}$, the so-called momentum compaction factor, relates the relative variation of path length to the relative variation of beam momentum $p$. It is worth mentioning that $\alpha_{\rm c}$ itself is a function of the particle's momentum offset, and in some applications this dependence must be taken into account. The value of the relativistic gamma at transition energy is defined as
\begin{equation}
\gamma_{\rm t} = \frac{1}{\sqrt{\alpha_{\rm c}}}
\end{equation}
and another important quantity is the so-called slip factor $\eta$ defined as
\begin{equation}
\eta = \frac{1}{\gamma_{\rm t}^2} -\frac{1}{\gamma^2} \, .
\end{equation}

The longitudinal beam dynamics are tightly linked with these three interconnected parameters, as the synchrotron tune $Q_{\rm s}$ is given by~\cite{sylee}
\begin{equation}
 Q_{\rm s} = \frac{\omega_{\rm s}}{\omega_{\rm rev}} = \sqrt{\frac{h \, e \, V_{\rm RF} |\eta \cos \phi_{\rm s}| }{2\pi \beta^2 E}}   \, ,
\end{equation} 
where $\omega_{\rm s}$ is the angular synchrotron frequency, $\omega_{\rm rev}$ the angular revolution frequency, $h$ the harmonic number, $V_{\rm RF}$ the RF voltage, $\phi_{\rm s}$ the phase of the synchronous particle, $\beta$ the relativistic factor, and $E$ the particle's energy.

Close to transition energy, i.e. for $\eta \to 0$, the adiabaticity condition~\cite{sylee}, given by
\begin{equation}
    \left \vert \frac{1}{\omega_{\rm s}^2} \frac{{\rm d} \omega_{\rm s}}{{\rm d} t} \right \vert \ll 1
\end{equation}
is violated and $Q_{\rm s} \to 0$. This is not the only adverse effect of the transition energy, as the longitudinal space-charge forces are also a potential source of disturbances given that their defocusing or focusing effect changes depending on whether the motion occurs below or above the transition energy. This means that a longitudinal mismatch is possible when the transition energy is crossed, which entails filamentation and eventually longitudinal emittance growth and beam losses. 

The remaining two critical aspects have to do with head-tail instabilities~\cite{inst1,inst2,inst3}, for which a chromaticity jump is imposed, as well as negative mass and microwave instabilities (see, e.g.~\cite{inst4} for a review of this aspect, and references therein). It is worth mentioning that intense experimental efforts have been devoted in recent years at CERN to study collective effects during  transition crossing in view of intensity upgrades~\cite{trstudies1,trstudies2,trstudies3} (see also~\cite{yannis} for a review).

As already mentioned, one of the techniques that has been developed to improve transition crossing is a so-called gamma-jump scheme (see, e.g.~\cite{gj1,gj2,gj3} and~\cite{gamjump} for reviews of the topic). The principle is based on the relationship between $\alpha_{\rm c}$ and optical parameters, namely
\begin{equation}
\alpha_{\rm c} = \oint \frac{D(s)}{\rho} \, d\, s \, ,
\end{equation}
where $D(s)$ is the dispersion function and $\rho$ is the local bending radius. Therefore, by manipulating the optical functions one can change $\alpha_{\rm c}$ and hence $\gamma_{\rm t}$. The idea is that while $\gamma$ is approaching the $\gamma_{\rm t}$ value corresponding to the nominal ring optics, the optics is modified so that the actual value of $\gamma_{\rm t}$ is pushed higher. Then, when approaching this new value, another very fast optics change brings the actual value of $\gamma_{\rm t}$ well below $\gamma$. An efficient gamma-jump scheme should ensure that a large enough $\Delta \gamma_{\rm t}$ is achieved, while at the same time the perturbation to the optical functions $\beta_{x,y}(s)$ and $D_{x}(s)$ is kept to a minimum. Additionally, the tune variation should be as small as possible for obvious reasons. These boundary conditions can, in general, be satisfied, but the price to pay is the need for a certain number of independently powered quadrupoles families with fast-pulsing capabilities. In the following it will be shown how all this can be simplified.

Parenthetically, it is worth mentioning that recently good progress has been made in the domain of beam dynamics of isochronous or quasi-isochronous rings (see, e.g.~\cite{iso1,iso2,iso3,iso4,iso5,iso6,iso7,iso8} and references therein) and in the use of the nonlinear dependence of $\alpha_{\rm c}$ on the momentum offset for the generation of the so-called $\alpha$-buckets~\cite{alpha1,alpha2,alpha3,alpha4,alpha5}. All this stands as a clear indication that non-linear effects are also being actively considered and used for designing novel applications in longitudinal beam dynamics.
\section{Stable islands: not only a means for transverse beam splitting}\label{sec:splitting}
The feature of stable islands that will be discussed here is not that related with beam trapping, i.e. with a dynamical variation of the position and size of the islands by means of a change of the tune, but rather the purely geometrical aspect linked to the existence of a fixed point located inside the stable islands. The concept that is crucial for the proposed method is that the presence of stable islands implies the existence of such fixed points, which are nothing else than a stable closed orbit in addition to that at the origin of the phase space, i.e. the nominal closed orbit. As already mentioned in the introduction, the periodicity of such an additional closed orbit is greater than one, meaning that this special orbit is longer than the ring circumference. 

The Poincar\'e section, i.e. the phase-space portrait at a given longitudinal location in the accelerator, of the horizontal betatron motion features ellipses as closed curves, whose area $2\pi J$ is an integral of motion. A symplectic transformation $(x,\ p)\ \rightarrow (\tilde{x},\ \tilde{p})$ maps the ellipse into a circle of radius $\sqrt{2J}=|z|$, where $z(s)=\tilde{x}(s)+i\tilde{p}(s)=\sqrt{2J}e^{-i\phi_x(s)}$, $\phi_x$ is the betatron phase advance, $2J$ is an invariant quantity, and  $(\tilde{x},\ \tilde{p})$ are referred to as Courant-Snyder (C-S) coordinates~\cite{courant}. Magnets generating nonlinear forces, such as sextupoles and octupoles, introduce a dependence of the betatron tune on the amplitude, in analogy with an anharmonic oscillator. At large amplitudes, the phase-space orbit is no longer an ellipse. Hence, it will no longer be mapped onto a circle by a C-S transformation. Nevertheless, it is possible to extend the approach, and the generalisation of the C-S coordinates in the presence of lattice non-linearities are usually referred to as Normal Forms coordinates~\cite{giallo} $\xi(s)=\sqrt{\rho}e^{-i\theta(s)}$. In Normal Forms, the phase space is partitioned by circles of radius $\sqrt{\rho}=|\xi|$ and, provided that $\rho$ is far enough from the separatrices, the radius is the new invariant. The Hamiltonian governing the dynamics can be written in action-angle variables $(\rho,\theta)$ as
\begin{equation}\label{eq:hamilton}
H=2\pi Q_x\rho -\frac{\Omega_2}{2}\rho^2 - R(\theta)\rho^2\ 
  +O(\rho^3)\ ,
\end{equation}
where $\Omega_2$ is the so-called amplitude-dependent detuning coefficient and $R(\theta)$ is a resonant term~\cite{diego} (not discussed here). The nonlinear tune $\nu_x$ can be written as 
\begin{equation}\label{eq:detuning}
\nu_x=\frac{1}{2\pi} \left \langle 
\frac{\partial H}{\partial \rho} \right \rangle_{\theta}
     = Q_x-\frac{\Omega_2}{2\pi}\rho\ -\frac{\langle R \rangle_{\theta}}{\pi}%
\rho\ ,
\end{equation}
where $\langle\ \rangle_{\theta}$ denotes the average over $\theta$. The detuning term $\Omega_2$ is given by~\cite{diego}
\begin{equation}
\Omega_2=-\frac{1}{16}\sum_l{K_{3,l}\beta_{x,l}^2} + 
	F(\beta_x,\phi_x, K_{2})\ ,
\end{equation}
where the first sum extends over all octupoles in the ring, and $\beta_x, \phi_x$ are the horizontal C-S parameters~\cite{courant}. The variables $K_{3}$ and $K_{2}$ are the integrated magnetic strengths of octupoles and sextupoles respectively, defined according to $K_n=\ell\,Z\,e/p \, (\partial^nB_y/\partial x^n) $, where $\ell$ represents the length of the magnet, $Z$ the charge state, $p$ the particle's momentum, and $B_y$ the vertical component of the magnetic field. 

The function $F$ contains the contribution of sextupoles and is represented by a polynomial of degree two, with monomials of the form $K_{2,i} \, K_{2,j}$ with $i, j$ running over the whole set of sextupolar elements. The coefficients are complex expressions of the beta functions at the locations $i$ and $j$ and of the phase advance between location $i$ and $j$ (the  explicit form may be found in~\cite{diego}). The control of $\Omega_2$ can be performed by using the strengths of octupole or sextupole magnets. 

The detuning with amplitude is the fundamental quantity for manipulating stable islands, since it allows the existence of fixed points when the linear tune $Q_x$ is moved away from the resonance condition $\overline{Q}_x$ by $\Delta=Q_x-\overline{Q}_x$. It can be shown that the fixed points are located at an amplitude
\begin{equation}\label{rho}
\rho^*(\Delta, \Omega_2)= -\frac{2\pi\Delta}{\Omega_2 + 
2 \, \langle R \rangle_\theta}
      \simeq-\frac{2\pi\Delta}{\Omega_2} + O(\Delta^2)\, .
\end{equation}
The remainder in the above equation includes the resonant term, which is proportional to $\Delta$~\cite{diego}. The weak linear dependence of $\Omega_2$ on $\Delta$, through $F$ and $\beta_x$, is here neglected, but it can be easily included in the computations if need be. The existence of fixed points is determined by the signs of $\Delta$ and $\Omega_2$. For a positive value of $\Omega_2$ fixed points will exist only when $Q_x< \overline{Q}_x$, $\rho^*$ being always positive. In this paper, $\Omega_2\ne 0$ will be assumed. If this condition is not met, higher-order terms need to be included, as done in~\cite{diego}. The theory allows the computation of the surface $\Sigma$ of the stable islands, as given by the area enclosed by the separatrices connecting the unstable fixed points~\cite{giallo}.

The last missing element to illustrate the idea behind the proposed scheme is the evaluation of the fixed point coordinates in the Cartesian frame $(x,p)$. It is worth noting that all coordinate transformations used so far, i.e. Courant-Snyder and Normal Forms, are always chosen to be tangential to the identity, thus implying that $\sqrt{\rho}e^{-i\theta}=\xi\simeq z + O(\rho)$. After some algebra, we obtain that 
\begin{eqnarray}
x^*(\Delta,\Omega_2,s)&\simeq&
	\sqrt{\beta_x(s)\, \rho^*(\Delta,\Omega_2)} \, \cos{\Phi(s)} 
	\label{x1}\\
p^*(\Delta,\Omega_2,s)&\simeq&-
	\sqrt{\frac{\rho^*(\Delta,\Omega_2)}{\beta_x(s)}} \times \nonumber \\
& \times & \Bigl[\alpha_x(s)\cos{\Phi(s)}+ \sin{\Phi(s)}\Bigr] \\  \label{p1}
\Phi(s)&\simeq&\phi_x(s)+\frac{2n\pi}{q}\ ,\quad 1 \leq n \leq q \, ,
\end{eqnarray}
which shows that the position of the stable fixed points can be precisely controlled by acting on the transverse tune and on the strength of the nonlinear magnets. The same holds true for the surface of the stable islands. 

\begin{figure}[htb]
\centering
  \includegraphics[trim=2truemm 0truemm 0truemm 2truemm, width=0.49\linewidth,angle=0,clip=]{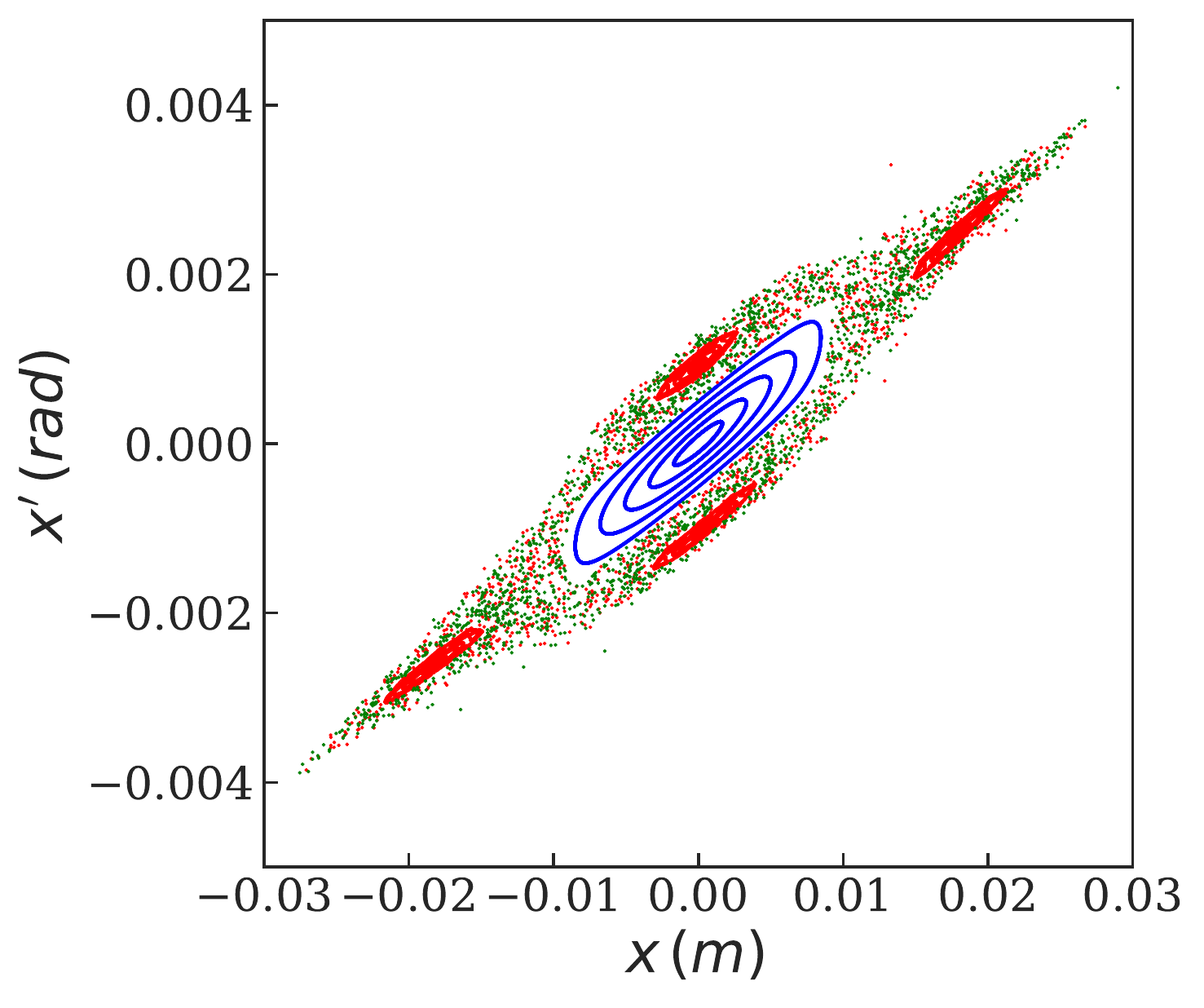}
  \includegraphics[trim=2truemm 0truemm 0truemm 2truemm, width=0.49\linewidth,angle=0,clip=]{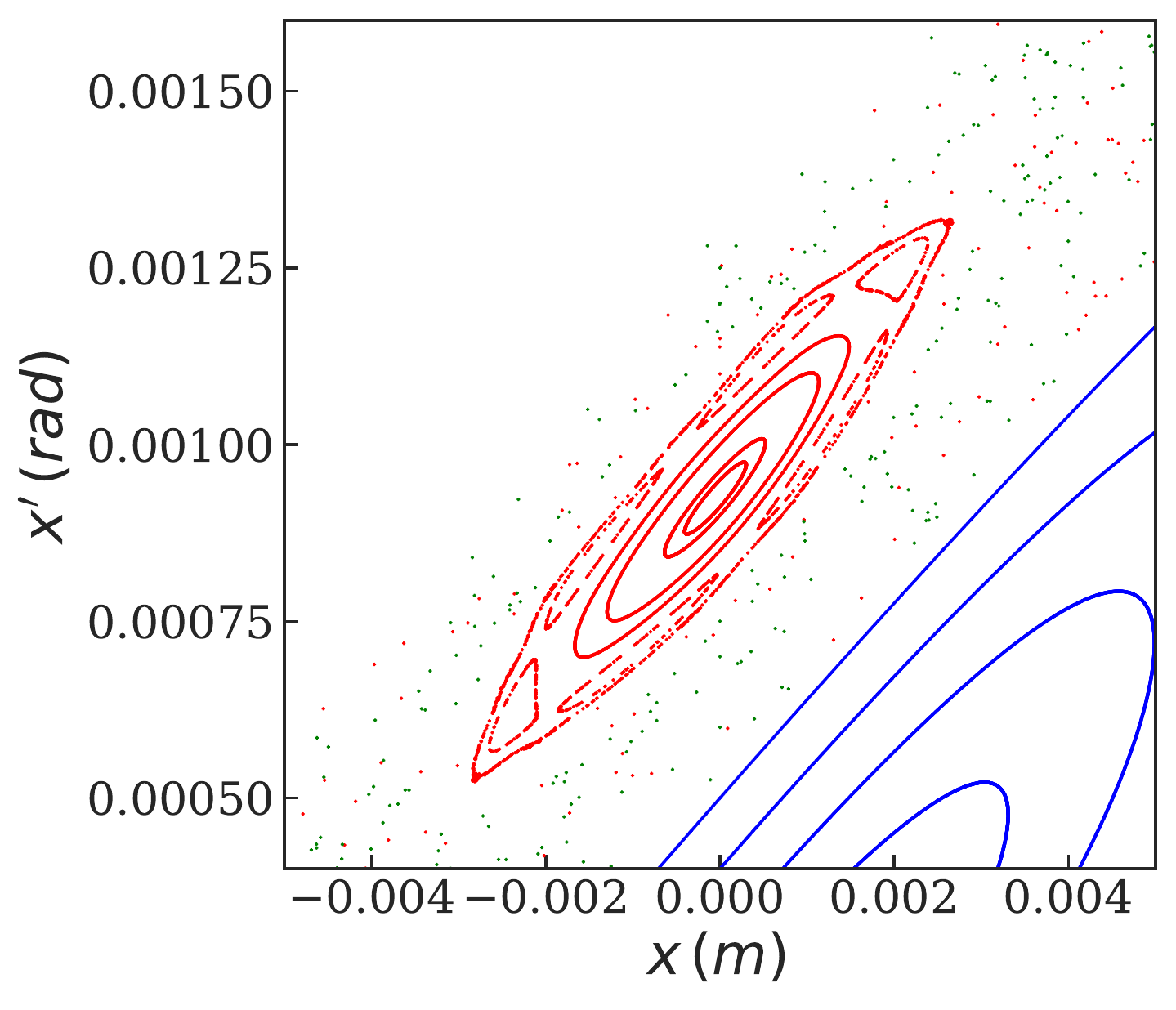}
  \caption{\label{phase_space} Left: Horizontal phase space of the lattice configuration used in later sections showing the presence of a chain of four stable islands. Right: zoom in around the stable island that is used in the proposed approach to cross transition. Note the chain of secondary stable islands inside the main stable island.}
\end{figure}

An example of the phase-space topology that can be created by using nonlinear magnets is shown in Fig.~\ref{phase_space} (left). A chain of four islands is visible, surrounding the closed orbit at the origin. In the right plot, a zoom of the phase space around the stable island that will be used for the proposed transition crossing manipulation is shown. The configuration of the lattice used to generate this phase space topology is presented in later sections.

\begin{figure}[htb]
\centering
  \includegraphics[trim=10truemm 35truemm 40truemm 5truemm, width=0.59\linewidth,angle=0,clip=]{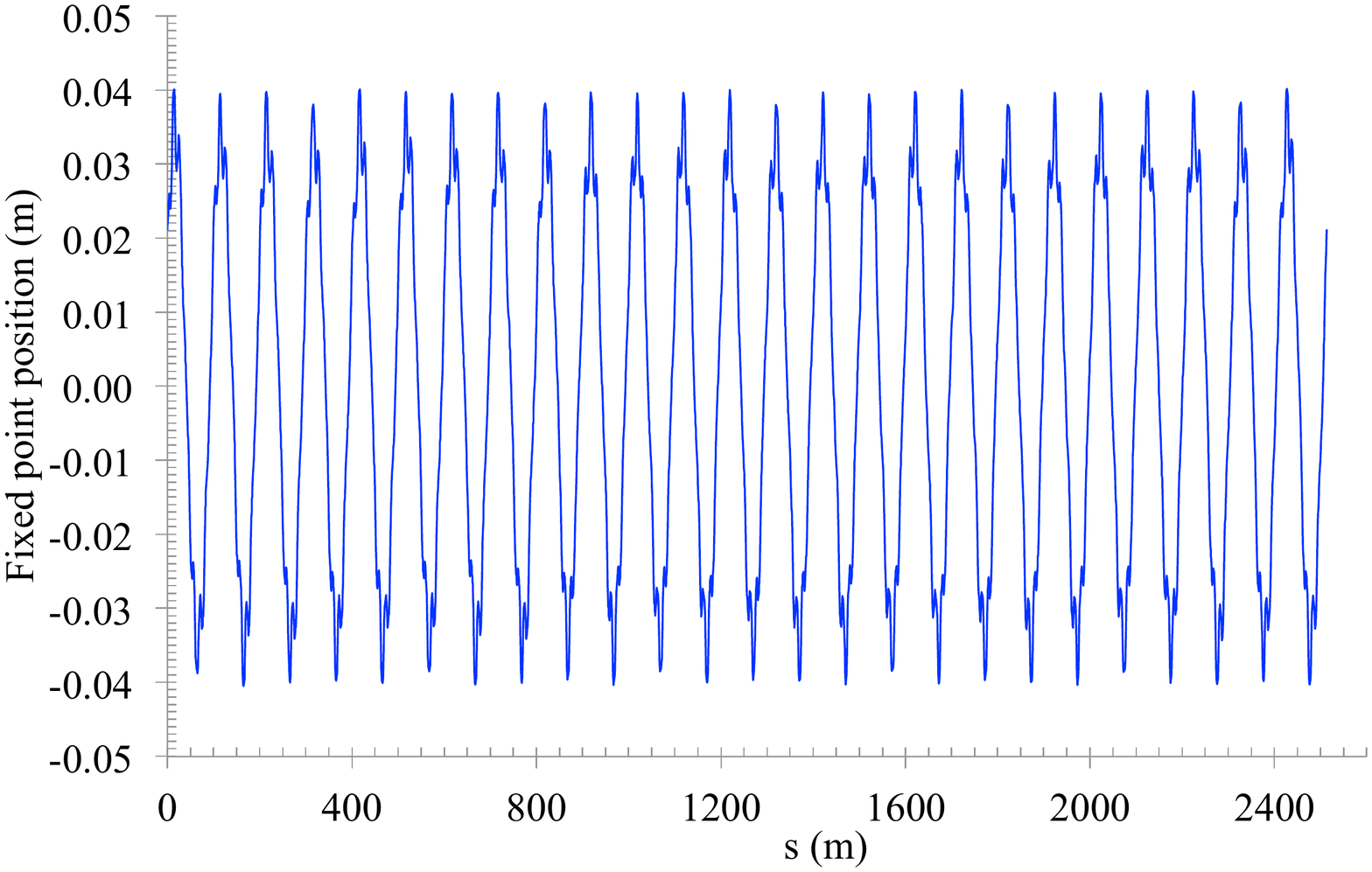}
  \caption{\label{fixed} Orbit at the fixed point linked with the fourth-order resonance as a function of $s$ along the PS ring.}
\end{figure}

\begin{figure}[htb]
\begin{tabular}{cc}
  \includegraphics[trim=5truemm 10truemm 10truemm 0truemm,width=0.48\linewidth,angle=0,clip=]{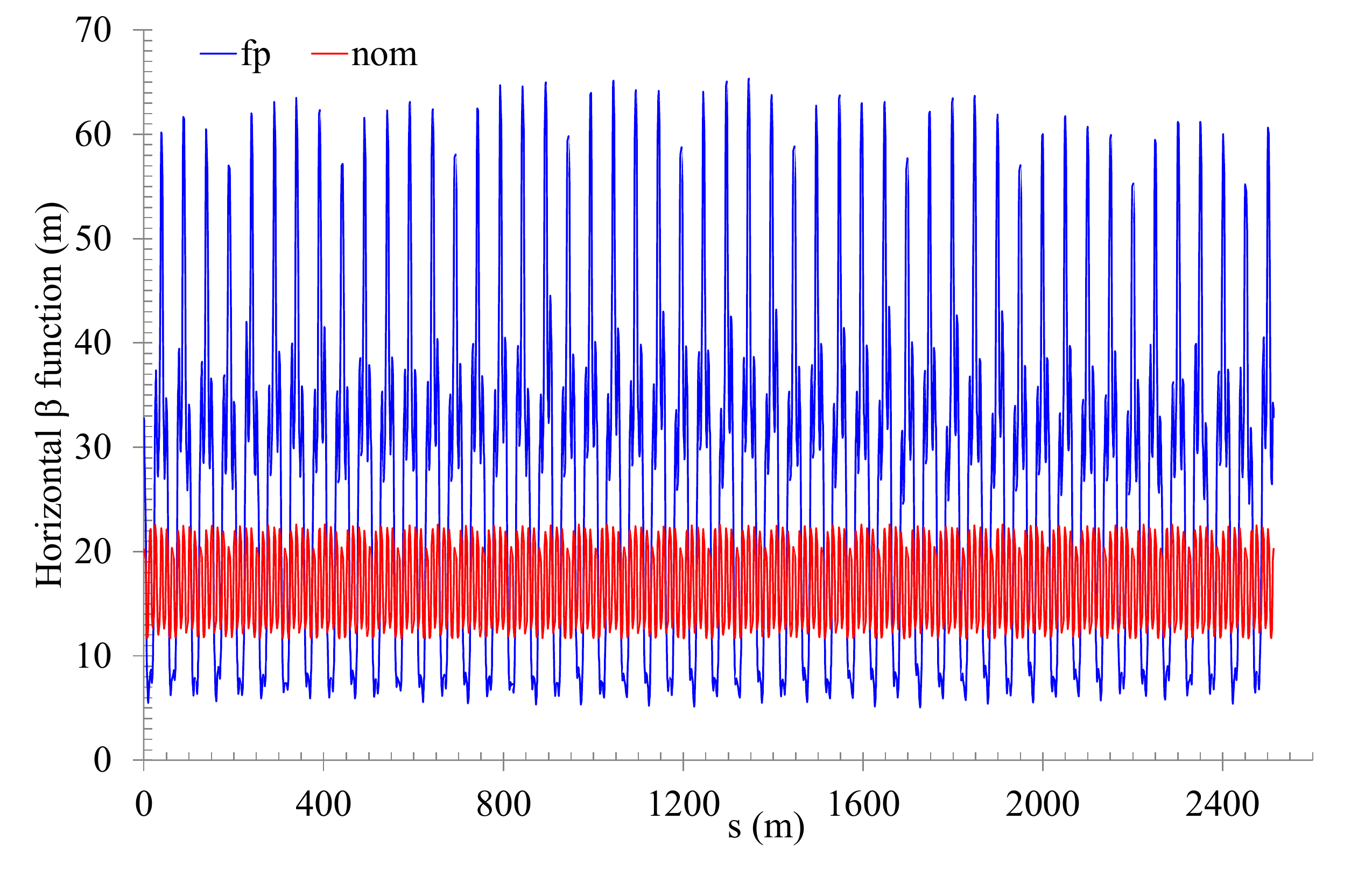} &
  \includegraphics[trim=5truemm 10truemm 10truemm 0truemm,width=0.48\linewidth,angle=0,clip=]{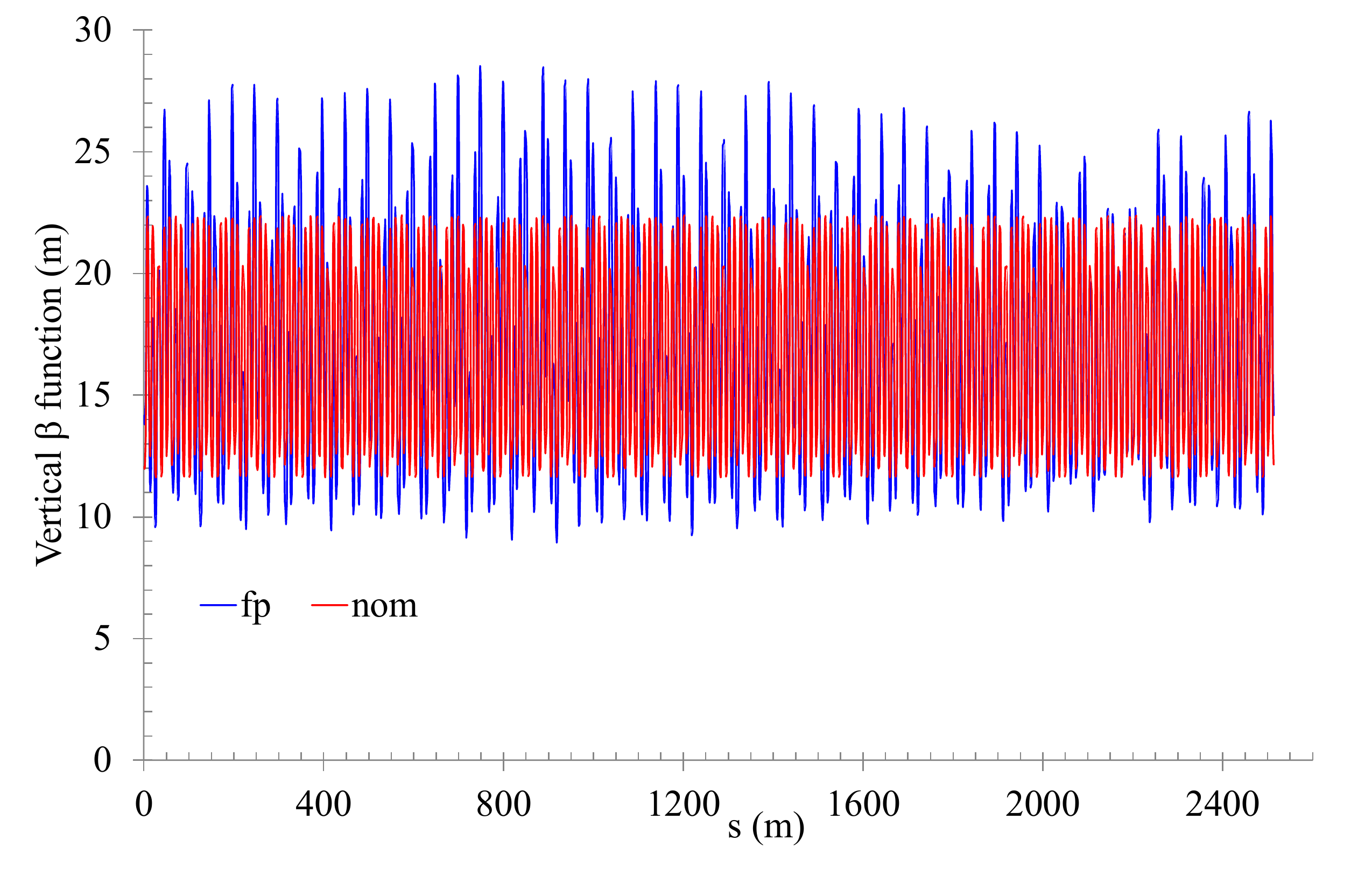}
\end{tabular}	
\caption{\label{optics} Horizontal (left) and vertical (right) beta function for the case of the beam dynamics around the nominal closed orbit (nom), and around the fixed points inside the islands (fp). The large difference of the beta values for the horizontal case is clearly visible.}
\end{figure}

\begin{figure}[htb]
\centering
  \includegraphics[trim=10truemm 0truemm 10truemm 0truemm,width=0.59\linewidth,angle=0,clip=]{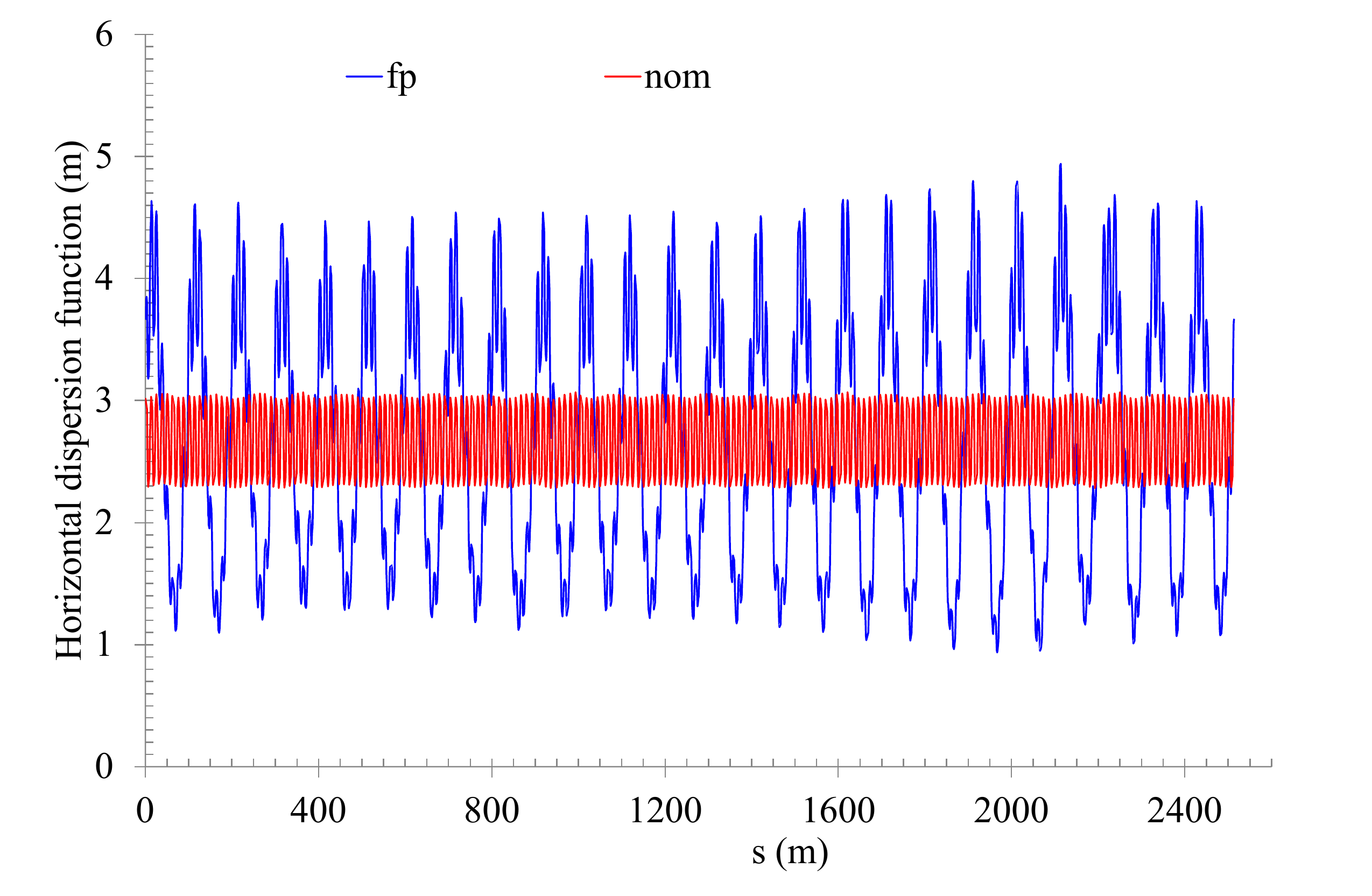}
  \caption{\label{disp} Horizontal dispersion function for the case of the dynamics around the nominal closed orbit (nom), and around the fixed points inside the islands (fp). The large difference of the dispersion values for the horizontal case is clearly visible.}
\end{figure}

The theory presented above allows the dynamics around the origin to be described as well as giving information about the existence and position of higher-order fixed points. However, if the dynamics around the fixed points has to be analysed, the original Hamiltonian should be expanded around the higher-order fixed points. In this case, the feed-down effect due to off-axis traversal of quadrupole and higher-order magnets has an impact on the motion of the particles, which will differ from that of a particle moving close to the origin. This implies that the ring optics for such an off-axis particle will be different, i.e. the parameters such as $\beta, \alpha$ and the horizontal dispersion will not be the same as the corresponding ones for the nominal closed orbit. An example of this is shown in Figs.~\ref{fixed},~\ref{optics},~\ref{disp}, where the orbit, $\beta$, and dispersion are shown for both the nominal closed orbit and the fixed point related with the fourth-order resonance. 
The model used for these computations is described in detail in the appendix, but the key point is that it is an effective model to represent the measured non-linearities of the $100$ main magnets of the PS ring. The computations were performed using MAD-X~\cite{MADX} in combination with the Polymorphic Tracking Code (PTC)~\cite{PTC}. For all cases, the $s$-variable spans four PS turns (the ring circumference is $200\pi$~m or about $630$~m). The orbit at the fixed point features large oscillations in the range $\pm 4$~cm, which is still compatible with the PS elliptical vacuum chamber geometry, whose typical half dimensions are $\pm 7$~cm (horizontally) and $\pm 3.5$~cm (vertically). 

The optical parameters are shown in Fig.~\ref{optics} for both the horizontal (left) and vertical (right) planes. The large difference between the nominal values of the horizontal $\beta$-function, i.e. the optical parameters for the beam dynamics close to the origin, and those for the motion inside the stable islands is clearly visible, with almost a factor of three for the peak-values of the $\beta$-function. The situation is different in the vertical plane, where although the $\beta$-values are not exactly the same, they are quite similar. This variation of the vertical $\beta$-values can be understood by considering that although no linear coupling is present in the model, the nonlinear dynamics introduces a coupling between the two planes, which is small but not completely negligible. 

Similar conclusions can be drawn from the behaviour of the dispersion functions for the two closed orbits as shown in Fig.~\ref{disp}. The direct comparison is striking, revealing a strong difference in beam dynamics between the motion close to the origin, i.e. around the nominal closed orbit, and that around the additional stable fixed points in the islands . 

As was mentioned earlier, the difference in optical parameters is due to the feed-down effects of higher-order fields impacting the beam dynamics. This implies that the optical parameters for the beam dynamics inside the islands are amplitude-dependent, i.e. if the fixed points are displaced the optics changes and, in particular, if the fixed points are moved towards the origin the optical parameters should tend to those of the nominal closed orbit. It is worthwhile stressing that by changing the distance of a particle from the fixed points inside the stable islands, the optical parameters it experiences will vary, even though the fixed point position is not changed. 

\begin{figure}[htb]
\centering
  \includegraphics[trim=5truemm 0truemm 10truemm 0truemm,width=0.48\linewidth,angle=0,clip=]{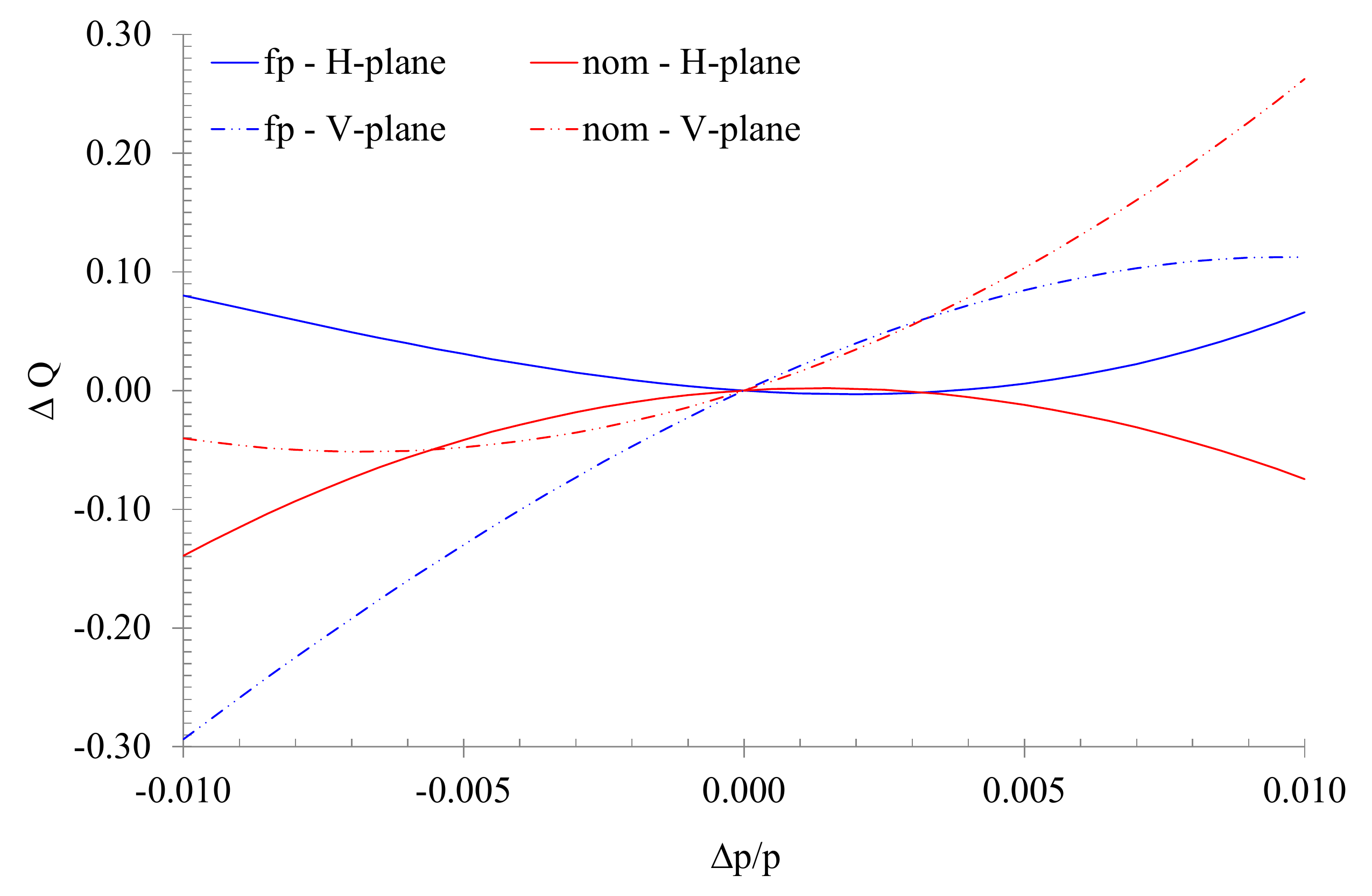}
  \includegraphics[trim=5truemm 0truemm 10truemm 0truemm,width=0.48\linewidth,angle=0,clip=]{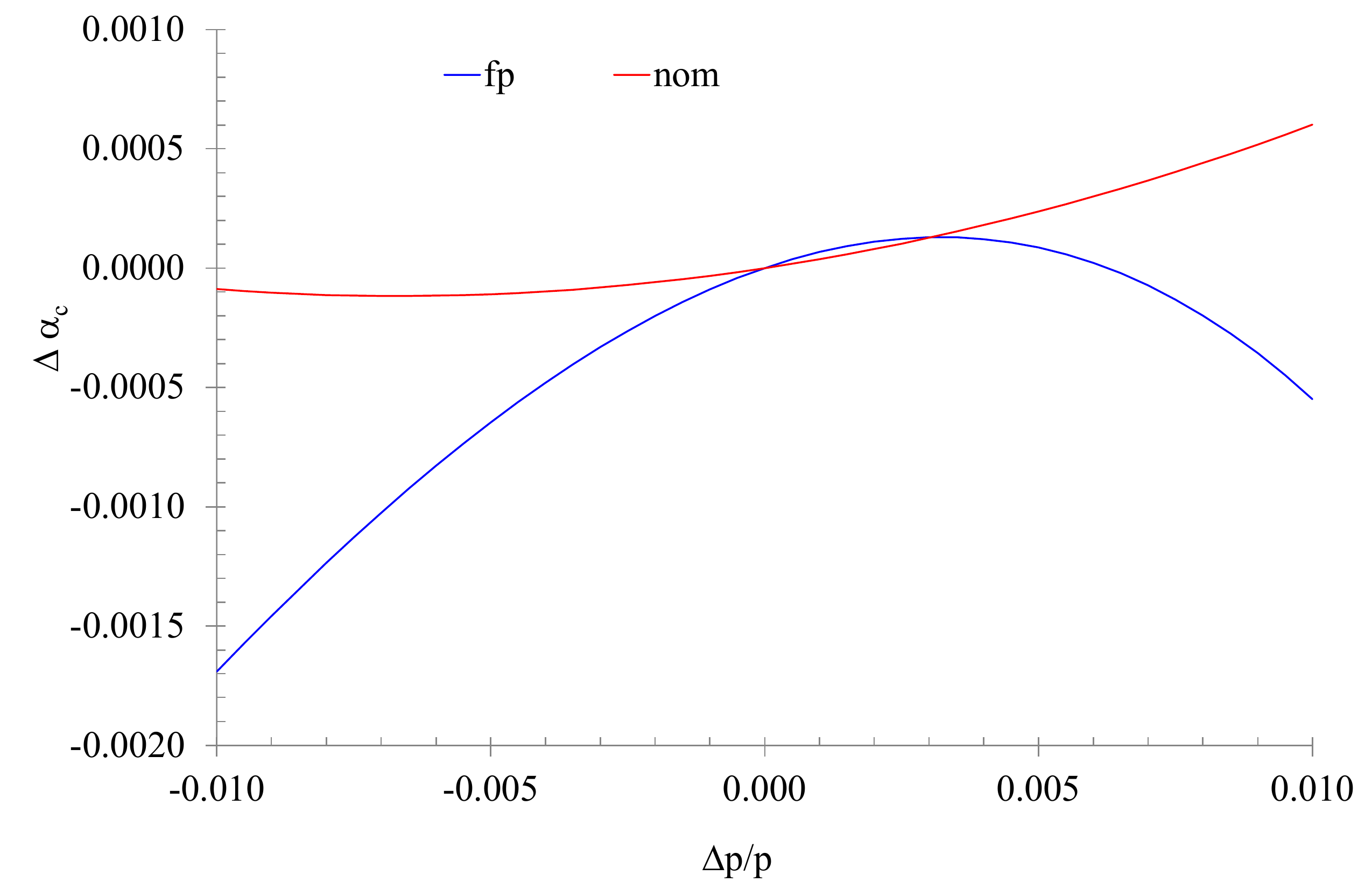}
  \caption{\label{chroma_acomp} Horizontal and vertical tune difference (left) and $\alpha_{\rm c}$ difference (right) as a function of momentum offset for the case of the beam dynamics around the nominal closed orbit (nom), and around the fixed points inside the islands (fp): the differences are clearly visible.}
\end{figure}

The difference in the dispersion functions for the two dynamics hints at a difference in the chromatic behaviour of the motion in the two cases, and this is clearly visible in Fig.~\ref{chroma_acomp} where the tune difference (left) and the $\alpha_{\rm c}$ difference (right) are plotted as a function of the momentum offset. 

Interestingly enough, the tune variation features different signs of the derivatives for the case of the fixed points inside the islands and the nominal closed orbit, thus showing that the linear chromaticity can have the opposite sign for the two closed orbits, and that this occurs in the horizontal as well as the vertical plane. The situation is similar for the difference of $\alpha_{\rm c}$ as a function of momentum offset where the curves referring to the two closed are concave and convex with opposite slopes at the origin. All this shows clearly how the the properties of the linear motion around the origin and the higher-order fixed points are different and can be acted upon by controlling the non-linear effects, which have no impact on the linear motion around the origin. 

The values of $\alpha_{\rm c}$, $\gamma_{\rm tr}$, and $E_{\rm tr}$ for the nominal orbit and the fixed points with zero momentum offset, as given by the numerical computations, are:
\begin{align}
\alpha_{\rm c, nom} & = 0.02223 & \alpha_{\rm c, fp} & = 0.02240 \nonumber \\
\gamma_{\rm tr, nom} & = 6.707332 & \gamma_{\rm tr, fp} & = 6.681233 \\
E_{\rm tr, nom} & = 7.148540 & E_{\rm tr, fp} & = 7.128724 \nonumber \, .
\end{align}
Although the difference between these values is small, of the order of $-0.78\%$, this is the key feature on which the new method is based.
\section{The proposed scheme}\label{sec:new_scheme}
\subsection{The principle}
As most of the elements required by the novel transition-crossing scheme have been presented in the previous sections, it just remains to clarify the details. 

Magnets generating nonlinear magnetic fields should be used to create stable islands with controlled properties, such as amplitude, surface, and $\alpha_{\rm c, island}$. In particular, the constraint is that
\begin{equation}
\alpha_{\rm c, fp} > \alpha_{\rm c, nom} \quad \Rightarrow \quad \gamma_{\rm t, fp} < \gamma_{\rm t, nom} \, .
\label{mechanism}
\end{equation}

\begin{figure}[htb]
\centering
  \includegraphics[trim=35truemm 30truemm 50truemm 25truemm,width=0.58\linewidth,angle=0,clip=]{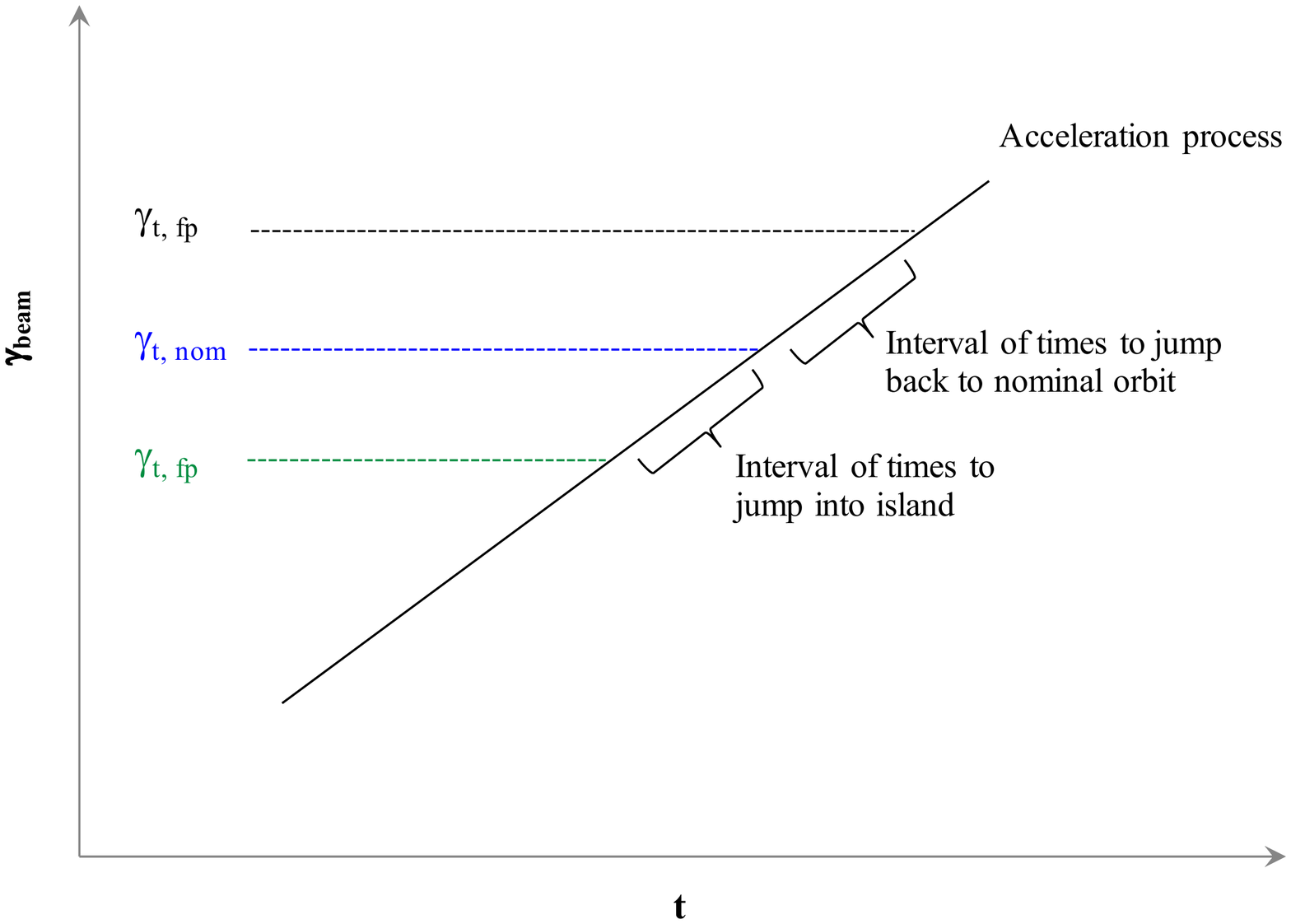}
  \caption{\label{fig:sketch} Sketch of the proposed transition-crossing process. The evolution of $\gamma_{\rm beam}$ as a function of time is shown together with the value of $\gamma_{\rm t, fp}$ (green) and $\gamma_{\rm t, nom}$ (blue) and the time interval for jumping into the island and then back to the nominal closed orbit. An alternative strategy is also given, with $\gamma_{\rm t, fp}$ (black) higher than than $\gamma_{\rm t, nom}$, which is equivalent to the one considered in our studies. The choice between the two options should be based on optics considerations.}
\end{figure}

The stable islands will be generated and available in the horizontal phase space, but not populated initially, as the beam will be circulating on the nominal closed orbit. During acceleration, when $\gamma \to \gamma_{\rm t, nom}$ a dipole kicker will be fired to deflect the beam to a stable island. Given the condition~(\ref{mechanism}), the beam in the islands will be far from the corresponding transition energy and its motion will not suffer. When the condition $\gamma > \gamma_{\rm t, nom}$ is fulfilled, i.e. the transition energy corresponding to the motion close to the origin has been crossed, the beam can be moved back to the standard closed orbit by means of another dipole kick. We remark that, strictly speaking, what is needed for the proposed approach to work is that $\gamma_{\rm t, nom}\neq \gamma_{\rm t, fp}$ and that their difference is large enough, as indicated by the sketch shown in Fig.~\ref{fig:sketch}.

In principle, generating the islands can be a static process, i.e. the fixed points could exist since injection. Alternatively, the nonlinear magnets might be slowly switched on to generate stable islands that are moving inwards, starting from high amplitudes, until they reach the required amplitude at the right time, i.e. when $\gamma \in [ \gamma_{\rm t, nom} - \varepsilon , \gamma_{\rm t, nom} + \varepsilon ] $ with $\varepsilon$ an arbitrary, positive quantity. It is evident that the surface of the islands should be large enough to accommodate the horizontal beam emittance without introducing any transverse emittance blow-up due to the nonlinear motion around the stable fixed point. 

The choice of the order of the resonance deserves some comment. It is a well-known fact that for a resonance of type $p/q, \,\,\, p,q \in \mathbb{N}$, the surface of the islands $\Sigma$ scales with the resonance order $q$~\cite{giallo}. To maintain a reasonable area this suggests the use of low-order resonances, and $q=4$ seems a good choice for the PS, for which there is also a good experience in terms of operational control thanks to MTE. Moreover, choosing an even value for $q$ has an appealing side effect. It has already been mentioned that the beam should be kicked twice during the process: the first time to move the beam to an island and then a second time to move it back to the origin. If the resonance order is even, then the phase space of the islands can be chosen such that at the kicker location two islands satisfy the condition $x^*=0$. Hence, for the same sign of the kicker deflection, one such island can receive the beam deflected from the origin, while the beam can be deflected back to the origin from the second island.

The advantage of the proposed scheme with respect to the classical gamma-jump scheme is that no pulsing magnets are needed except for the kicker, meaning that the beam dynamics around the origin of phase space is not perturbed as the proposed method relies on the use of non-linear beam dynamics, which by definition does not affect the linear motion around the origin. Last but not least, this method does not suffer from issues related to a jump of the tune, which affects all transition-crossing schemes based on fast optics changes. In fact, the tune corresponding to the motion around the origin needs to be close to the resonant value, at the level of $10^{-3}$ units. Hence, the tune change due to the jump between the two sets of closed orbits is completely negligible.
\subsection{The ring model}
\begin{figure}[htb]
\centering
  \includegraphics[trim=5truemm 0truemm 10truemm 0truemm,width=0.48\linewidth,angle=0,clip=]{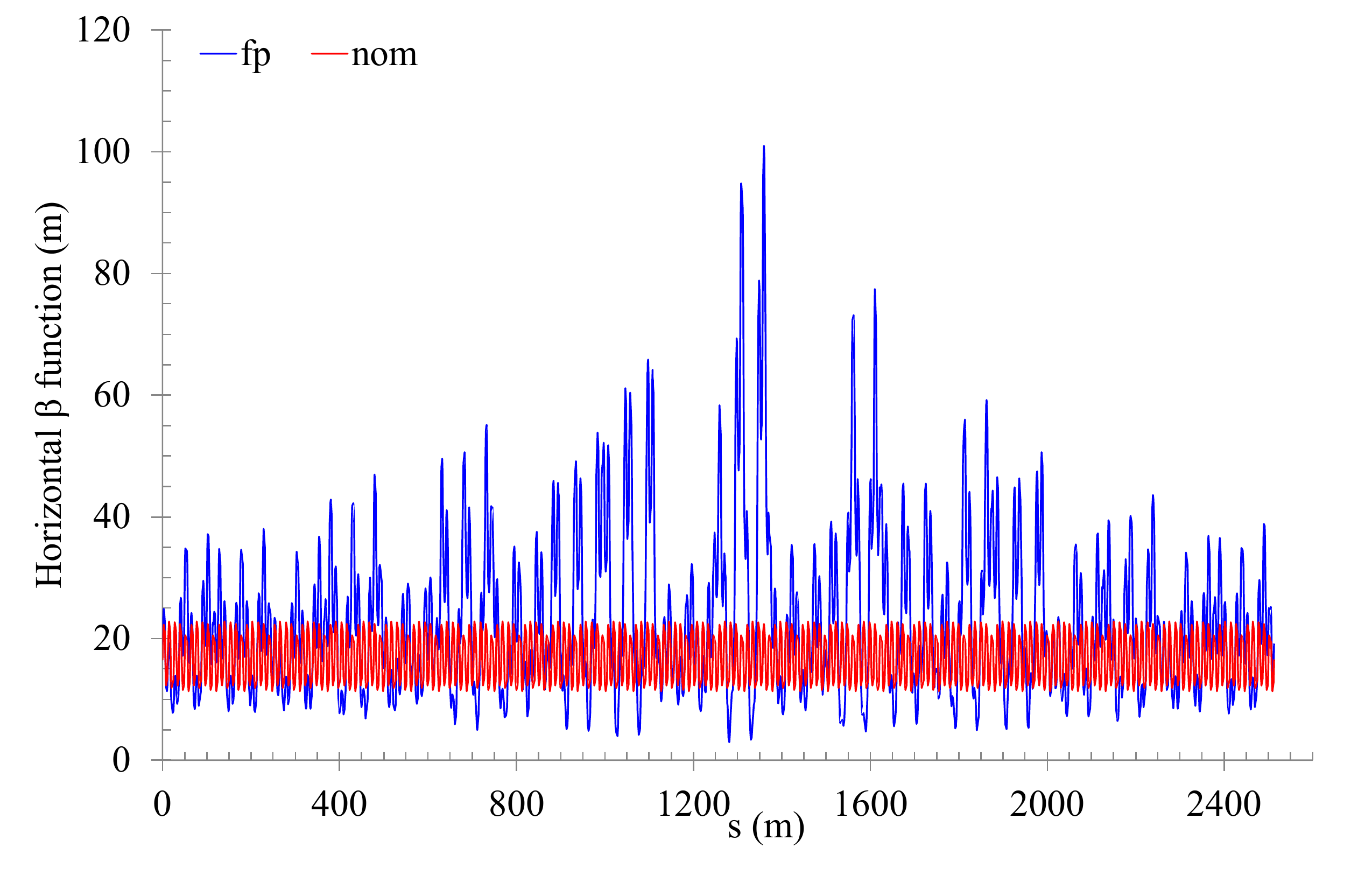}
  \includegraphics[trim=5truemm 0truemm 10truemm 0truemm,width=0.48\linewidth,angle=0,clip=]{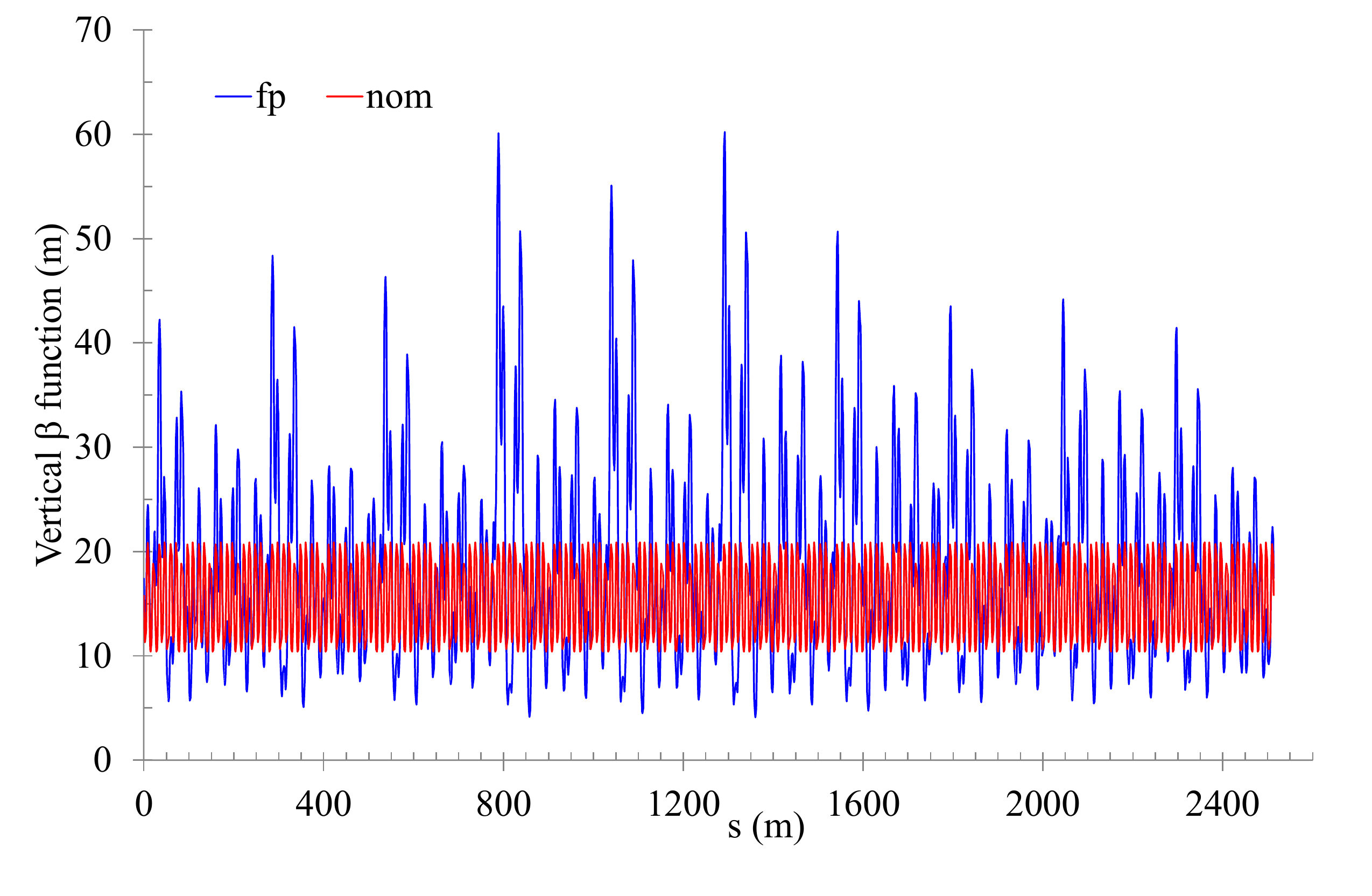}
  \includegraphics[trim=5truemm 0truemm 10truemm 0truemm,width=0.48\linewidth,angle=0,clip=]{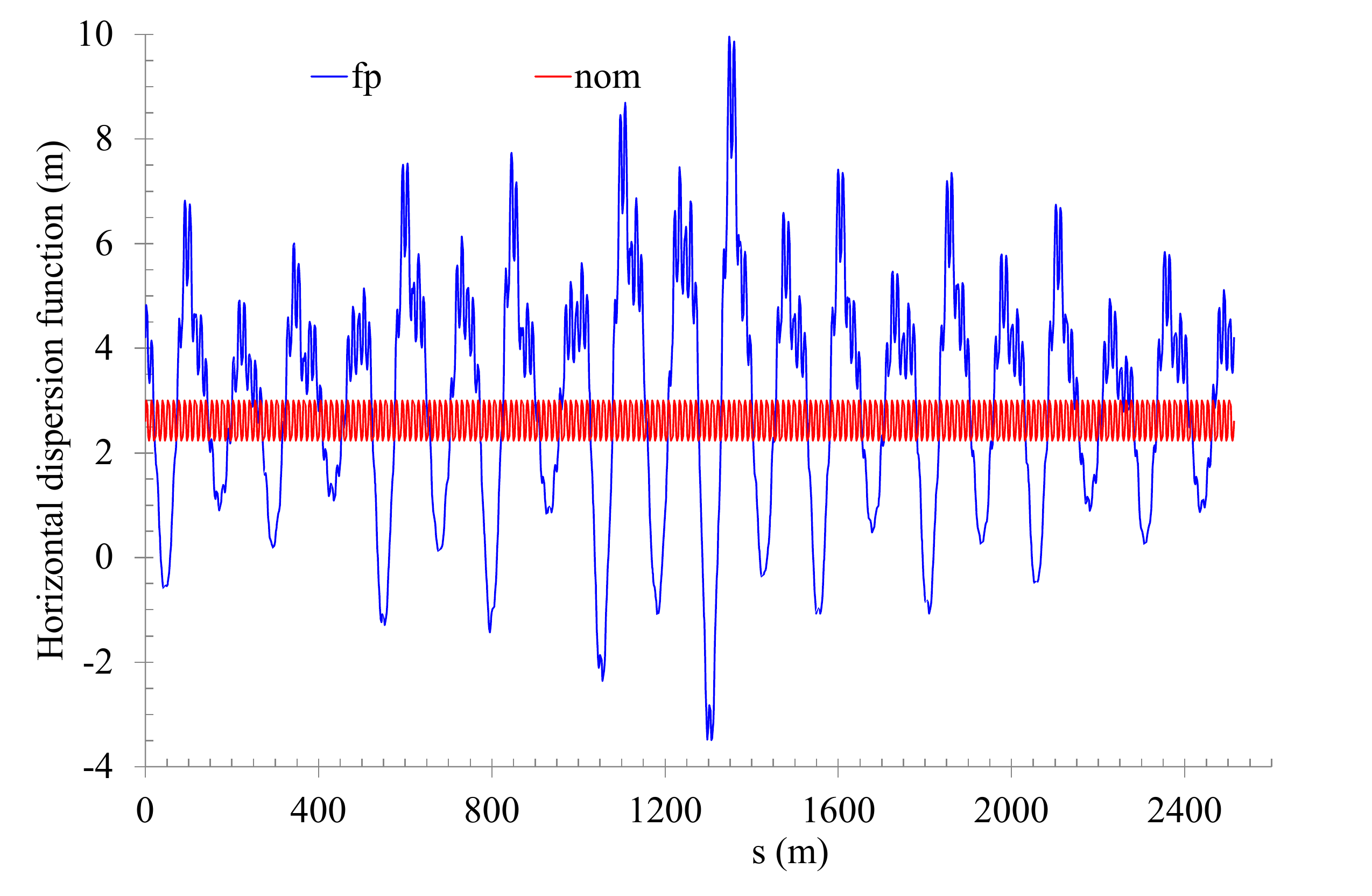}
  \includegraphics[trim=5truemm 0truemm 10truemm 0truemm,width=0.48\linewidth,angle=0,clip=]{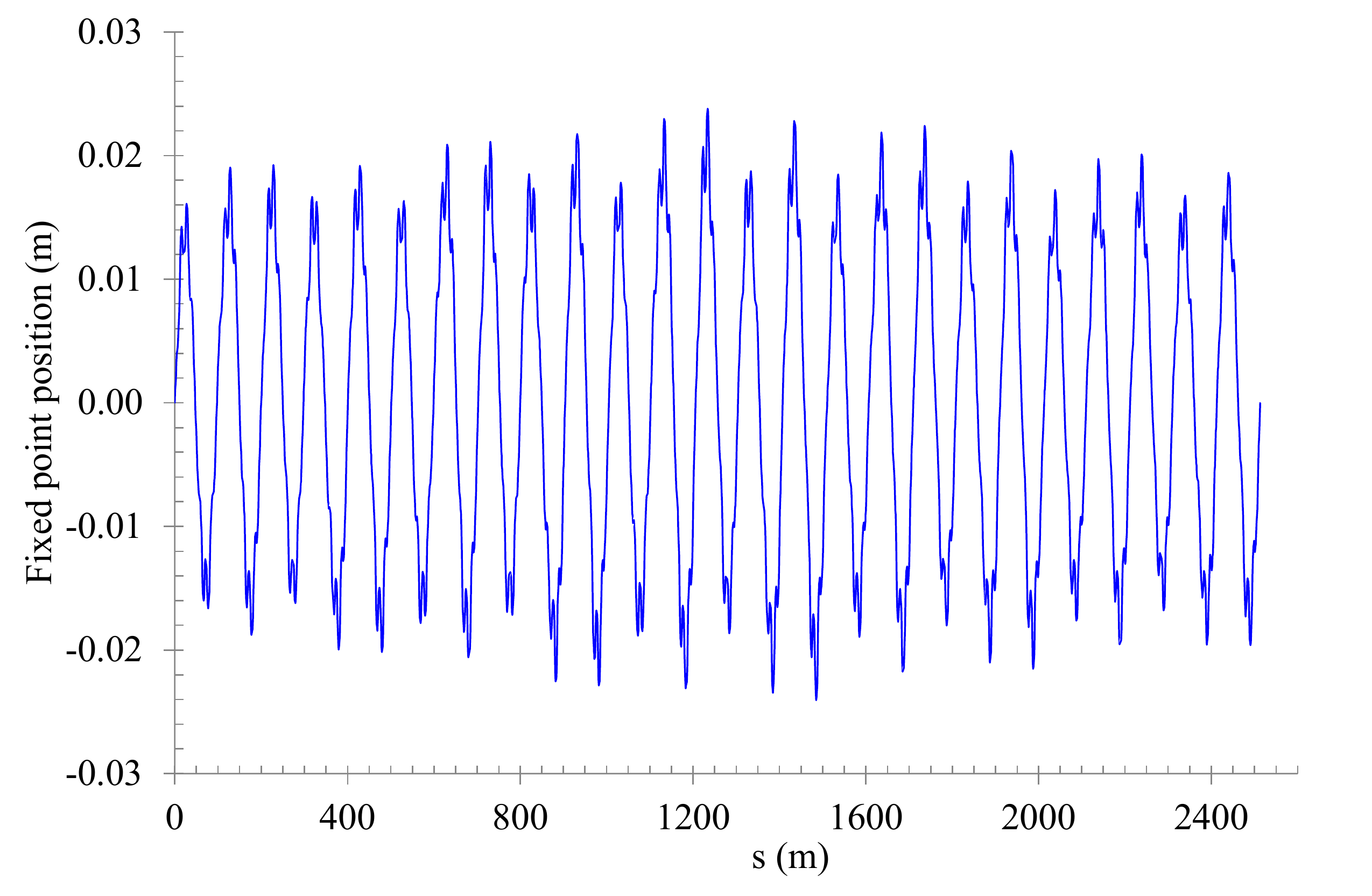}
  \caption{Evolution of the optical parameters for the nominal closed orbit (nom) and the fixed points inside the islands (fp) for the special PS lattice developed for a proof-of-principle simulation of the proposed approach to cross transition. The horizontal beta-function, the vertical beta-function, the horizontal dispersion, and the fixed point position are shown (left to right, top to bottom).}
\label{fig:model}
\end{figure}

To observe the features of this new technique proposed to cross the transition more clearly, a dedicated PS lattice has been prepared in view of performing proof-of-principle simulations. Non-linear components were added to the main dipoles in order to enhance the difference between $\gamma_{\rm t, fp}$ and $\gamma_{\rm t, nom}$, and the selected configuration features $\gamma_{\rm t, fp}=5.58$ and $\gamma_{\rm t, nom}=6.18$. Furthermore, individual sextupoles and octupoles have been installed in selected straight sections to generate the appropriate non-linear effects. In Fig.~\ref{fig:model} the main optical parameters are shown for the nominal closed orbit and the fixed points inside the islands.

The differences in optical parameters for the nominal closed orbit and the fixed points inside the islands are clearly visible. This is certainly expected in the horizontal plane, where the proposed manipulation occurs. However, as the sources of non-linearities are not always located where $\beta_{y}$ is small, the vertical dynamics is also perturbed. This is indicated by the difference between $\beta_{y}$ for the nominal closed orbit and the fixed points inside the islands (as already seen in Fig.~\ref{optics}). The horizontal dispersion is also strongly perturbed, which is explained by the fact that the non-linear effects are not stemming only from locations with a small dispersion value. All in all, these observations indicate that the prepared lattice is certainly not the ideal configuration to implement this novel approach to transition crossing. Nevertheless, this lattice is certainly suitable for a proof-of-principle study. 

Another key point is the search for the optimal location of the kicker that is used to deflect the beam into the islands and then back to the nominal closed orbit once the transition energy has been crossed. Two main criteria have been used: firstly, at the kicker location there should be at least one island whose centre lies at $x=0$ to allow deflection of the beam to the centre of the island, thus minimising the emittance blow up; secondly, the mismatch between the optical parameters of the beam on the nominal closed orbit and those at the fixed points inside the islands should be minimised to avoid emittance growth due to the sudden displacement of the beam from the nominal closed orbit to the island and vice versa. To quantify this effect, we applied the following indicator~\cite{bryant}
\begin{equation}
 H = \frac{1}{2} \left [ \frac{\beta_{\rm nom}}{\beta_{\rm fp}} + \left ( \alpha_{\rm nom} - \alpha_{\rm fp} \frac{\beta_{\rm nom}}{\beta_{\rm fp}} \right )^2 \frac{\beta_{\rm fp}}{\beta_{\rm nom}} + \frac{\beta_{\rm fp}}{\beta_{\rm nom}} \right ] \, , 
\end{equation}
This is normally used to evaluate the impact of betatron mismatch at injection in a circular machine, and where $\alpha_{\rm nom}$ and $\alpha_{\rm fp}$ in this case indicate the corresponding Twiss parameter for the nominal closed orbit and the fixed point, respectively. The location found in this way was checked a posteriori for dispersion mismatch which was found to be in the order of $25$~\%. 

Note that to simplify the search for the optimal location of the kicker, it has been assumed that it can provide both a positive and a negative deflection. In this way a single device can be used to control the passage between the nominal closed orbit and the fixed points inside the islands. It is worth mentioning that, with enough flexibility in the design of the lattice, one could prepare a configuration of the islands such that at a given location the two (top and bottom) islands have centres at $x=0$: in this way, with a single-sign deflection, hence a simplified hardware design of the kicker, one could deflect the beam from the nominal closed orbit to the top island and then, after an appropriate number of turns, when the beam in the top island has rotated in phase space to be in the bottom island, the bottom island can be kicked back to the nominal closed orbit.   
\subsection{Results of numerical simulations}
The 6D numerical simulations were carried out using the code PyORBIT~\cite{pyorbit}. An energy ramp was simulated assuming an RF voltage $V_{\rm RF}$ ranging from $0.2$~MV to $2$~MV, a stable phase of $30^\circ$ degrees, and total energy  $E_{\rm t}$ from $4.4$~GeV to $8$~GeV. To limit the CPU time of the simulation to a few tens of hours, the whole transition-crossing process was performed in $9\times 10^3$ turns. 

\begin{figure}[p]
\centering
    \begin{tabular}{ccc}
  \multicolumn{3}{c}{\includegraphics[trim=0truemm 0truemm 0truemm 0truemm,width=\linewidth,angle=0,clip=]{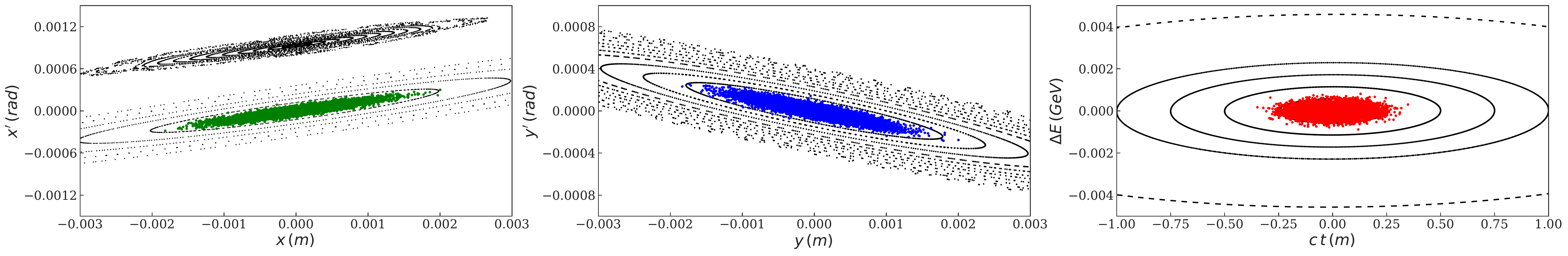}} \\
  \hspace{8truemm} \includegraphics[trim=0truemm 0truemm 0truemm 0truemm,width=0.27\linewidth,angle=0,clip=]{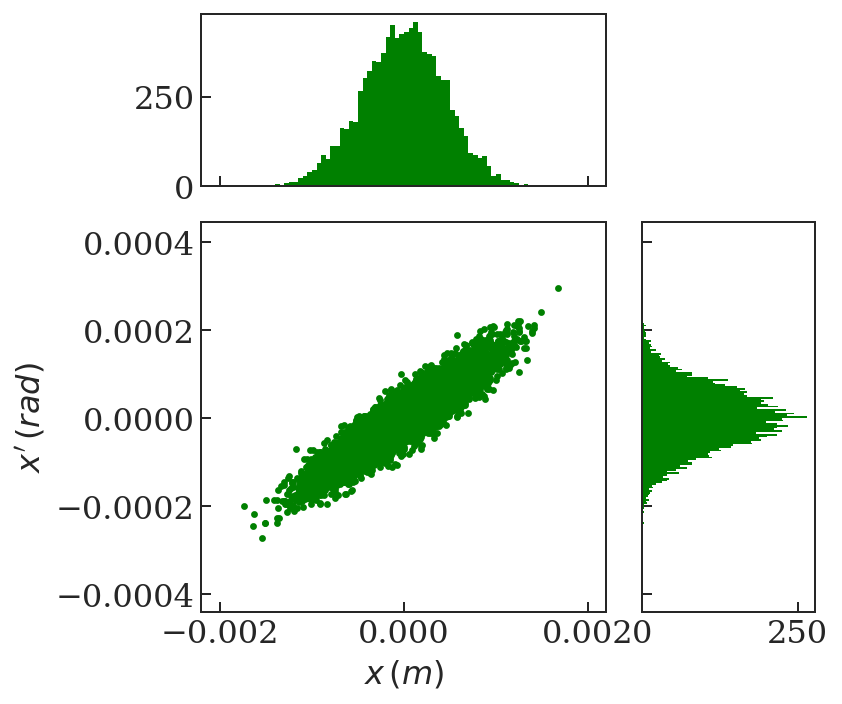} &
  \hspace{4truemm} \includegraphics[trim=0truemm 0truemm 0truemm 0truemm,width=0.27\linewidth,angle=0,clip=]{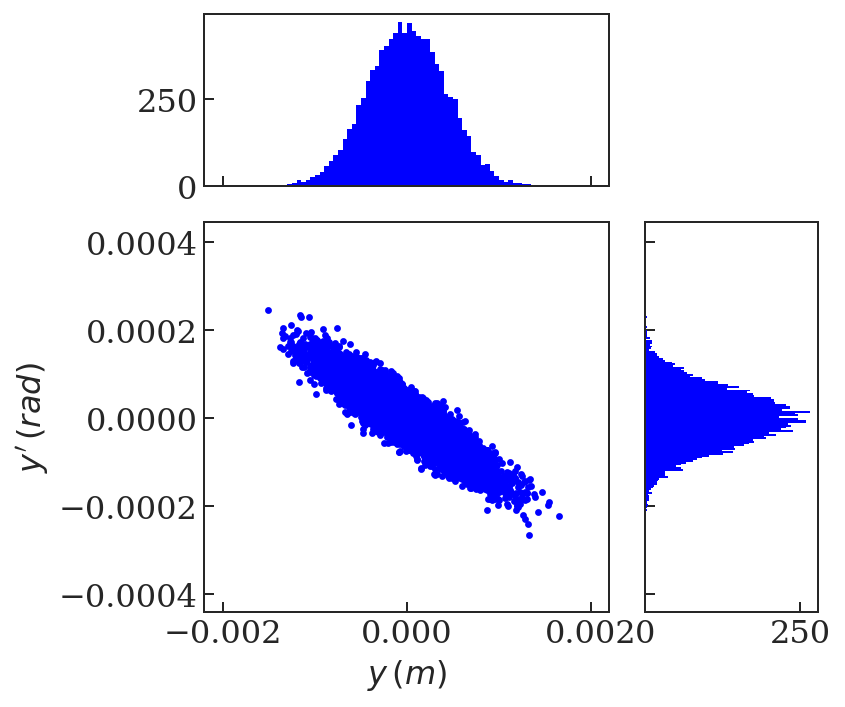} &
  \hspace{0truemm} \includegraphics[trim=0truemm 0truemm 0truemm 0truemm,width=0.27\linewidth,angle=0,clip=]{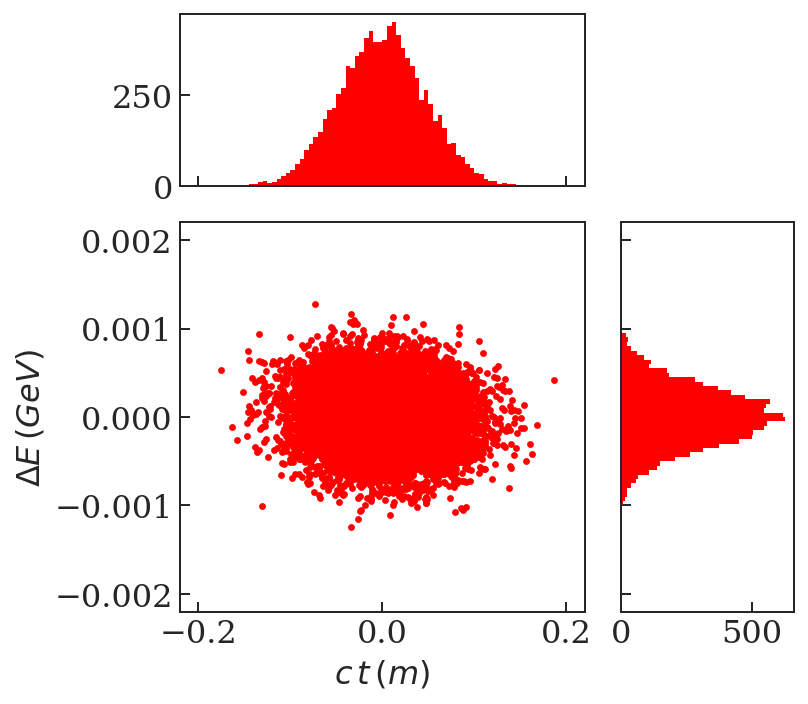} \\
  \multicolumn{3}{c}{\includegraphics[trim=0truemm 0truemm 0truemm 0truemm,width=\linewidth,angle=0,clip=]{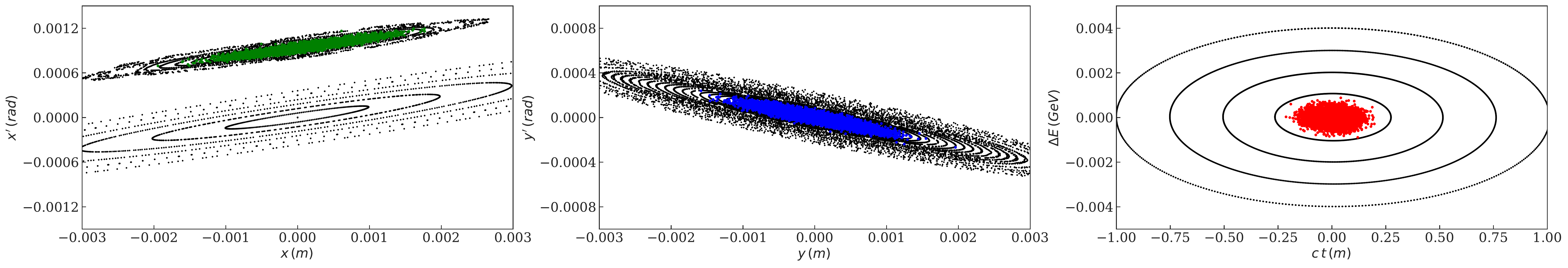}} \\
  \hspace{8truemm} \includegraphics[trim=0truemm 0truemm 0truemm 0truemm,width=0.27\linewidth,angle=0,clip=]{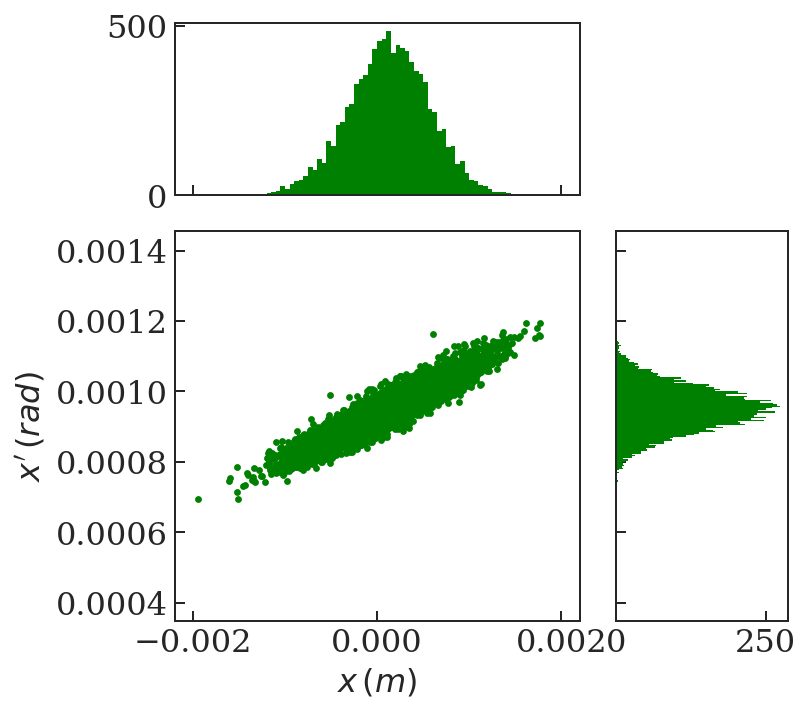} &
  \hspace{4truemm} \includegraphics[trim=0truemm 0truemm 0truemm 0truemm,width=0.27\linewidth,angle=0,clip=]{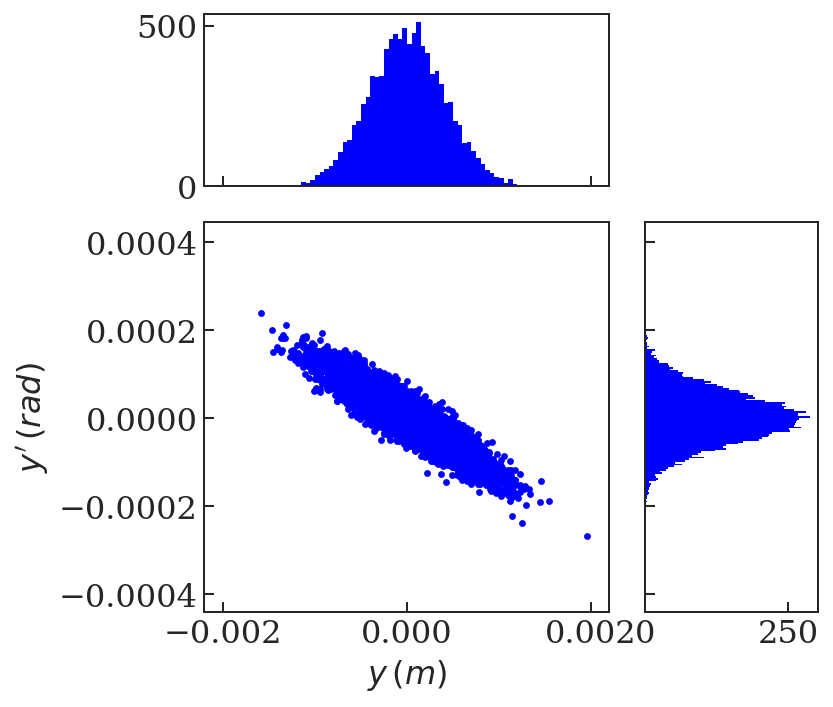} &
  \hspace{0truemm} \includegraphics[trim=0truemm 0truemm 0truemm 0truemm,width=0.27\linewidth,angle=0,clip=]{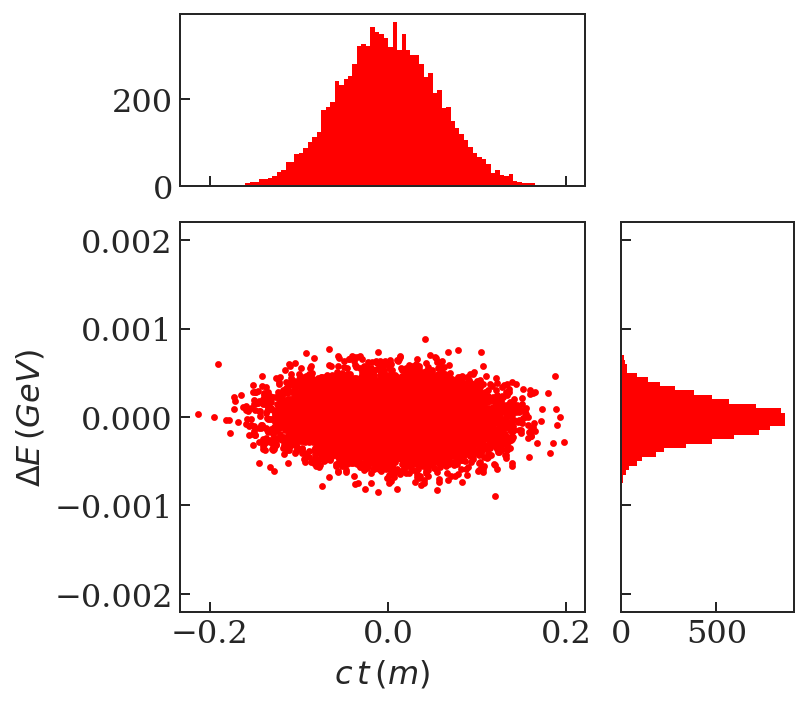} \\
  \multicolumn{3}{c}{\includegraphics[trim=0truemm 0truemm 0truemm 0truemm,width=\linewidth,angle=0,clip=]{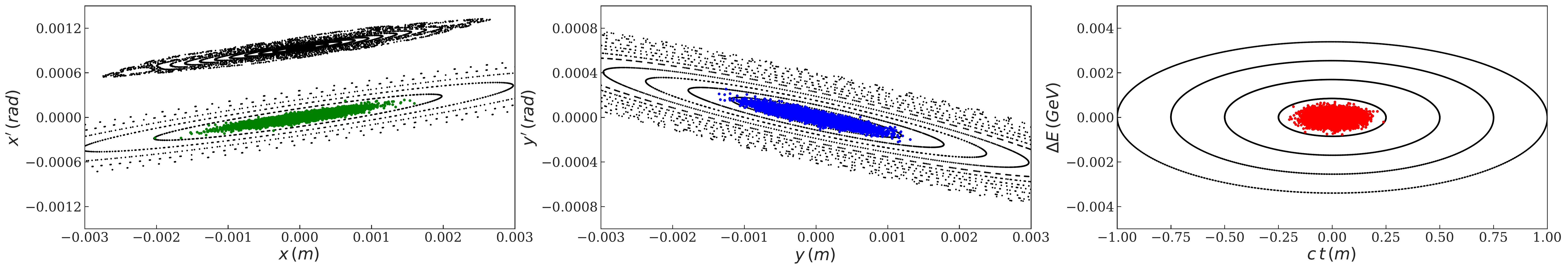}} \\
  \hspace{8truemm} \includegraphics[trim=0truemm 0truemm 0truemm 0truemm,width=0.27\linewidth,angle=0,clip=]{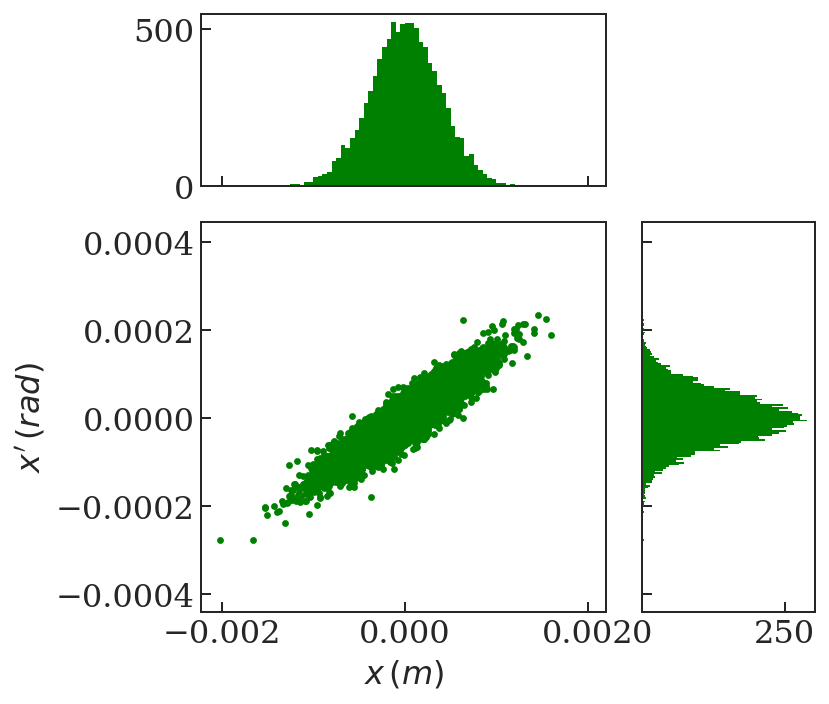} &
  \hspace{4truemm} \includegraphics[trim=0truemm 0truemm 0truemm 0truemm,width=0.27\linewidth,angle=0,clip=]{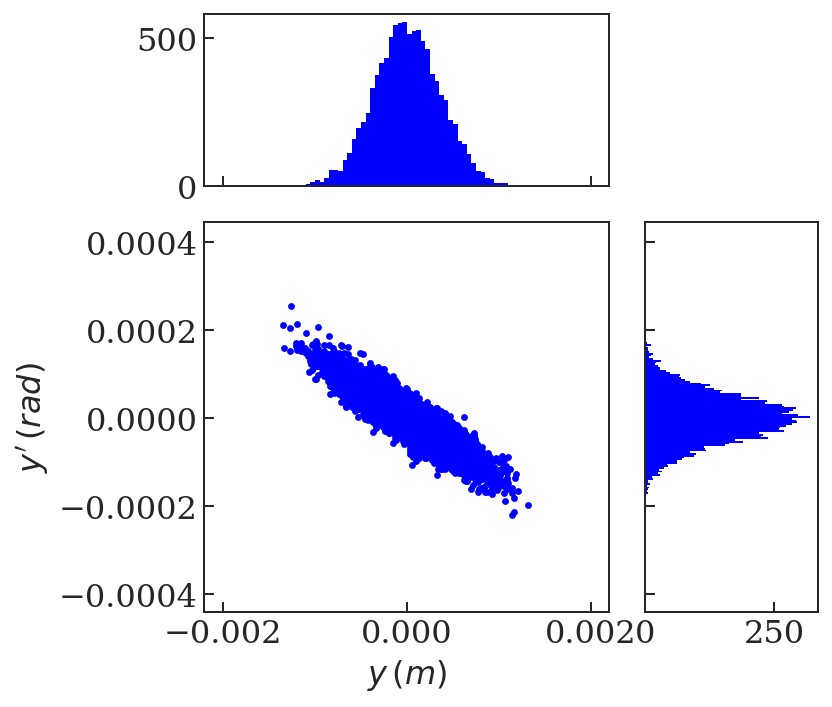} &
  \hspace{0truemm} \includegraphics[trim=0truemm 0truemm 0truemm 0truemm,width=0.27\linewidth,angle=0,clip=]{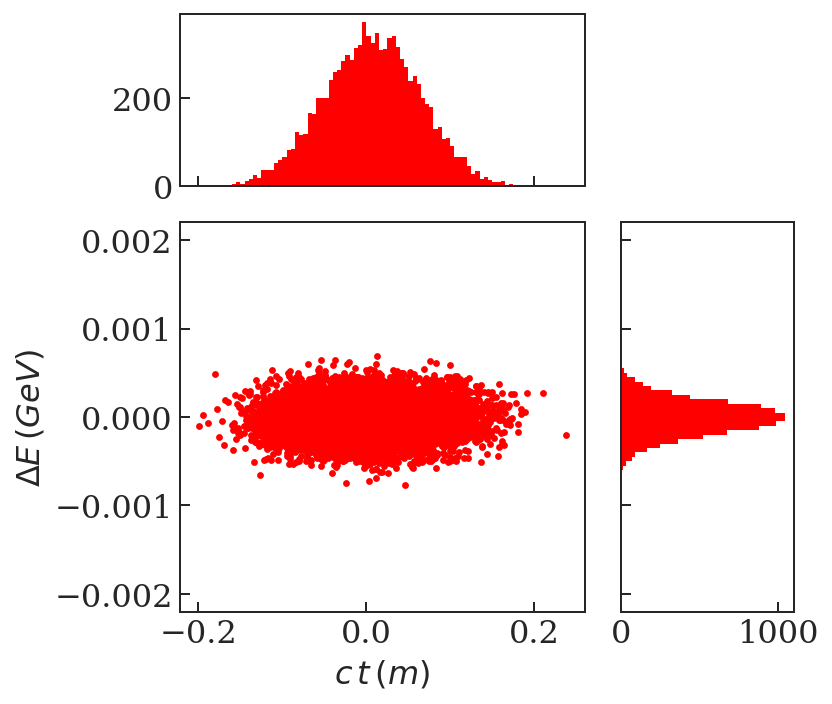} \\
    \end{tabular}
  \caption{Phase-space portraits and beam distributions in 6D for the various stages of the beam acceleration through transition (horizontal, vertical, and longitudinal planes, left to right). Top row: initial beam distribution on the nominal closed orbit. Middle row: beam distribution in the island just before the kick back to the nominal closed orbit. Bottom row: final beam distribution after completion of the whole transition-crossing process. The case depicted here refers to $V_{\rm RF}=0.8$~MV, and to the following normalised rms emittance values: $\epsilon^*_{H}=\epsilon^*_{V}=68.7 \times 10^{-3}~\mu$m, and $\epsilon_{L}=2.1 \times 10^{-4}$~eVs.}
\label{fig:simul1}
\end{figure}

A Gaussian distribution in 6D is generated at $E_{\rm t} =4.4$~GeV and it is accelerated on the nominal closed orbit over $3\times 10^3$ turns up to $E_{\rm t}=5.6$~GeV (note that $\gamma_{\rm t, nom}$ corresponds to $E_{t}=5.8$~GeV)\footnote{Note that $\gamma_{\rm t, fp}$ corresponds to $E_{t}=5.2$~GeV}. This stage is mainly used to let the initial distribution filament and adapt to the phase-space geometry of the nominal closed orbit. Then, the beam is displaced in the island and accelerated there for $3\times 10^3$ turns up to $E_{\rm t}=6.8$~GeV. At this stage, the transition energy has been crossed and the beam is kicked back to the nominal closed orbit and accelerated over $3\times 10^3$ turns up to $E_{\rm t}=8$~GeV. The outcome of these simulations is shown in Fig.~\ref{fig:simul1} for one of several cases considered in this study. 

The beam distributions are overlaid with phase-space portraits to highlight the phase-space topology. The first observation is that the surface of the stable island determines the maximum value of the normalised rms horizontal beam emittance, as an emittance larger than the island surface would lead to beam losses, which are completely absent in the simulations presented here. The distortion of the beam distribution due to the proposed manipulations is essentially invisible on this scale. The numerical computations reveal a rather small emittance growth, corresponding to $< 4$~\% and $< 2$~\%, for the horizontal and vertical normalised emittances respectively. Another observation is that some non-linear coupling between the horizontal and vertical planes is visible in the phase-space portraits shown in the middle row of Fig.~\ref{fig:simul1}. In fact, the vertical phase-space portrait shown in the middle row is not the same as the phase-space portraits shown in the top row and bottom row, and the only difference is the type of initial conditions in the horizontal phase space, which are around the nominal closed orbit (top row and bottom row) or around the fixed point in the island (middle row). 

\begin{figure}[htb]
\centering
  \includegraphics[trim=2truemm 0truemm 2truemm 0truemm,width=0.32\linewidth,angle=0,clip=]{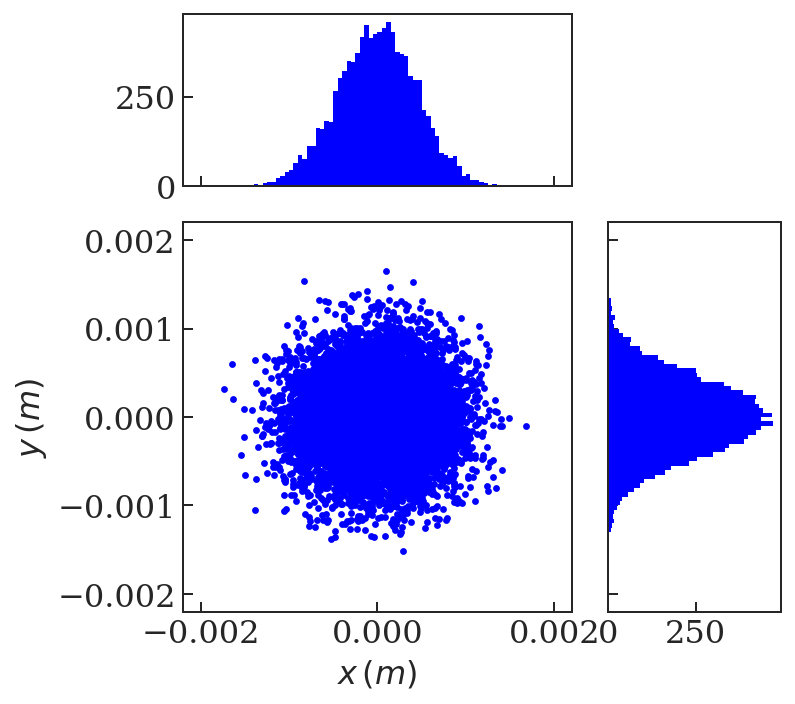}
  \includegraphics[trim=2truemm 0truemm 2truemm 0truemm,width=0.32\linewidth,angle=0,clip=]{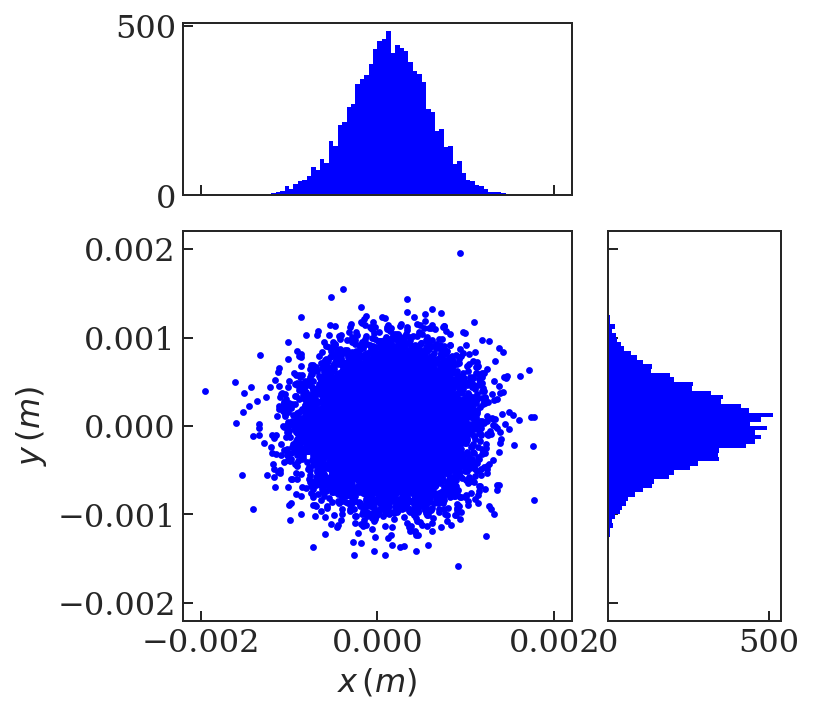}
  \includegraphics[trim=2truemm 0truemm 2truemm 0truemm,width=0.32\linewidth,angle=0,clip=]{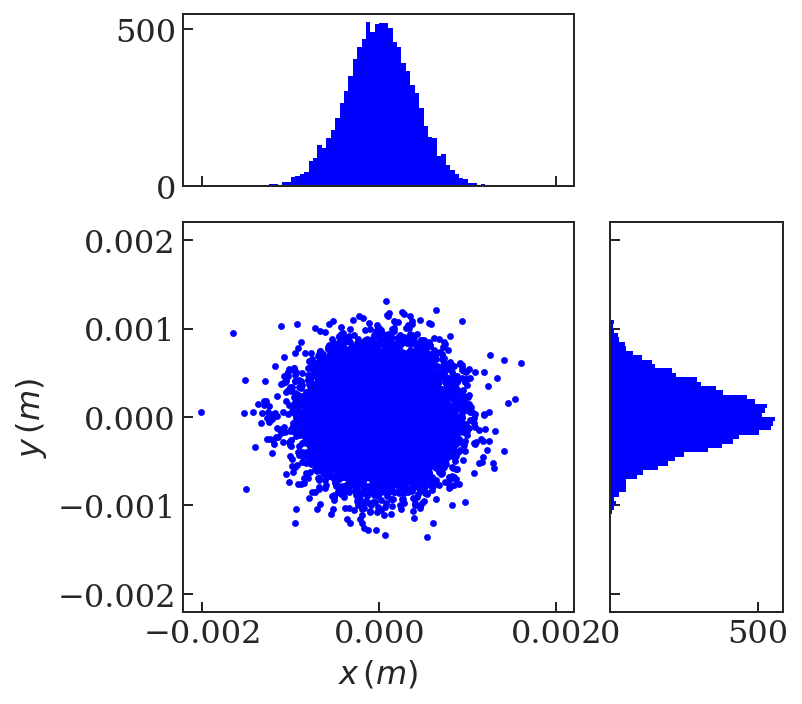}
  \caption{Physical-space beam distributions for the various stages of the beam acceleration through transition. Left: initial beam distribution on the nominal closed orbit. Middle: beam distribution in the island just before the kick back to the nominal closed orbit. Right: final beam distribution after completion of the whole acceleration process. The case depicted here refers to $V_{\rm RF}=0.8$~MV, and to the following normalised rms emittance values: $\epsilon^*_{H}=\epsilon^*_{V}=68.7 \times 10^{-3}~\mu$m, and $\epsilon_{L}=2.1 \times 10^{-4}$~eVs.}
\label{fig:simul2}
\end{figure}

The beam distribution in the physical space corresponding to the three stages of the proposed transition crossing strategy are shown in Fig.~\ref{fig:simul2}, where no major difference is observed across all stages of the transition crossing process.

The study included a series of detailed scans of the longitudinal and transverse emittances to inspect their growth. The beam distribution was initially assumed to be Gaussian in the horizontal plane and $\delta$-like in the  vertical and longitudinal planes. In this configuration, the growth of $\epsilon^*_{\rm H}$ is $< 4$~\% and occurs at the stage when the beam is kicked into the island. The beam distribution was then assumed to be Gaussian in both horizontal and vertical planes, and $\delta$-like in the longitudinal plane. A scan was then performed using the same value for the horizontal and vertical normalised beam emittance and their growth was found to be $< 7$~\% and $\approx 3$~\% respectively. In this case, some emittance growth also occurs in the stage when the beam is kicked back to the nominal closed orbit and the total growth shares equally between the stage in which the beam is kicked to the island and when it is pushed back to the nominal closed orbit. When the longitudinal emittance is varied, the case with maximum value of $\epsilon_{\rm L}$ is the one shown in Fig.~\ref{fig:simul1}, for which the values of the emittance growth have already been quoted. In all cases, the emittance growth is rather small, although it should be stressed that the longitudinal emittance is fairly small. The origin of the limitation for the maximum value of the longitudinal emittance comes from the chromatic effects related to the dynamics in the island, which could be cured by using a more flexible lattice.
\section{Discussion of the proposed approach}\label{sec:discussion}
The numerical simulations carried out confirmed the validity of the proposed method to perform the crossing of the transition energy. Still, some effects have been observed, which will be discussed briefly here. These are linked more with the model prepared for these numerical tests than with the proposed beam manipulation itself. 

Firstly, the surface of the stable islands can be controlled at will, as long as there are enough independent non-linear magnets available in the lattice. In fact, an analytical estimate of the island surface is available~\cite{giallo} and can be used at the design stage of the proposed manipulation to ensure flexibility in the value of the horizontal emittance that can be accommodated. 

Secondly, some non-linear coupling between the horizontal and vertical planes has been observed. This is due to the non-optimal location of the non-linear magnets used to create the stable islands and control their momentum compaction. Indeed, it is well known from theory and the experience gained with MTE (both from numerical simulations and the operational beam~\cite{MTEope1,MTEope2}) that placing the non-linear magnets in sections where $\beta_{y}$ is smaller than $\beta_{x}$ is a very effective strategy to make non-linear coupling negligible. 

Thirdly, the chromatic aberrations created by the non-linear magnets are also linked to their non-optimal placement. Lattice insertions with zero-dispersion function would allow a better control of island properties, as non-linear magnets located in a position with zero dispersion would not contribute to chromatic aberrations. By properly tuning the strengths of the magnets in zero and non-zero dispersion regions, it would be possible to have a better control of both the chromatic and non-chromatic effects, as some of the magnets would act selectively only on one of the two types of effects. This would allow the limitation on the value of the longitudinal emittance to be removed.

For all these reasons, it is believed that the effects observed would be easily controlled in circular accelerators featuring slightly more complex lattices than that of the PS.
\section{Conclusions}\label{sec:conclusions}
In this paper, a novel method to cross the transition energy in a circular hadron particle accelerator has been presented and discussed in detail. The principle is based on non-linear beam dynamics, which is used to generate stable islands in horizontal phase space. It has been shown that by carefully choosing the strength of the non-linear magnets, it is possible to select the optical properties of the fixed points of these islands such that the momentum compaction differs from that of the orbit at the origin of the phase space. This makes it possible to avoid crossing the transition energy by simply displacing the beam to a stable fixed point, and then back to the origin once the energy is sufficiently far from its transition value. This method does not rely any pulsing quadrupole, as for standard gamma-jump schemes, and hence leaves the linear beam dynamics around the origin unperturbed. Similarly, the issues due to the tune jump associated with standard transition-crossing schemes are not applicable to the novel method proposed.

Numerical simulations were performed as a proof-of-principle of this novel beam manipulation. The model of the standard PS lattice was modified to generate stable islands with a transition energy different enough from that of the normal closed orbit to apply the proposed method. No particle losses were observed during the whole beam manipulation process and very little emittance growth, at the level of $10$~\% or less, was observed in the transverse planes. All this was achieved with a beam distribution featuring a small longitudinal emittance, as the chromatic effects in the special PS lattice model used were non-negligible. This, however, is not a drawback that is intrinsic to the proposed method, but rather a limitation of the model used, which is based on the existing PS ring structure. Indeed, the chromatic effects can be perfectly controlled in a ring in which non-linear elements can be installed at correctly chosen locations to mitigate the chromatic effects and non-linear coupling between the two transverse planes. We therefore consider that the proposed method is indeed a suitable alternative to the standard techniques devised so far to overcome issues encountered during transition crossing in a particle accelerator.
\section*{Data availability}
Data sharing not applicable to this article as no datasets were generated or analysed during the current study.
\clearpage
\appendix
\section{The PS Ring}\label{sec:appendix}
The PS lattice (see also~\cite{MTE-prog,PS50V1,PSDR} for more details) consists of ten super-periods each made of ten combined function dipole magnets 4.4 m long, interlaced with eight 1.6 m and two 3.0 m straight sections. Every magnet is composed of two half-units with gradients of opposite sign, separated by a central junction. The latest magnetic measurements using Hall probes~\cite{PS50V1} showed that stray fields at the magnet ends introduce an additional quadrupolar component, while in the gap between the two half units a sextupolar component was observed. The fine adjustment of tunes and chromaticities is performed by means of extra coils mounted on the pole faces of the main dipoles, the so-called pole-face windings (PFW). Until 2007, this system was controlled by three independent currents, sufficient to set both tunes and the horizontal chromaticity. In 2008 an upgraded version was commissioned, able to control independently five physical parameters (such as tunes, chromaticities plus one additional free parameter) by means of five new separate circuits~\cite{Burnet}.

\begin{figure}[htb]
\begin{tabular}{cc}
  \includegraphics[width=8.5cm,angle=0,clip=]{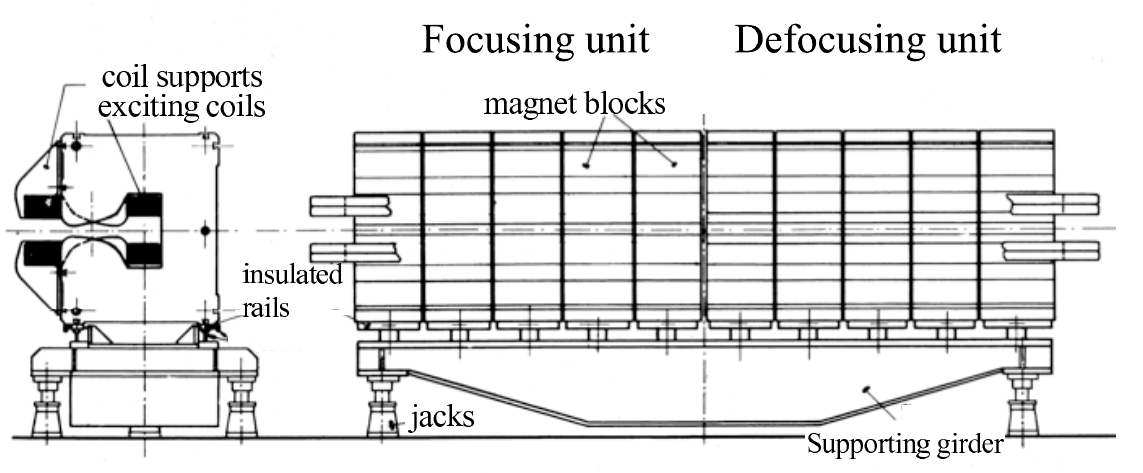} &
  \includegraphics[width=8.5cm,angle=0,clip=]{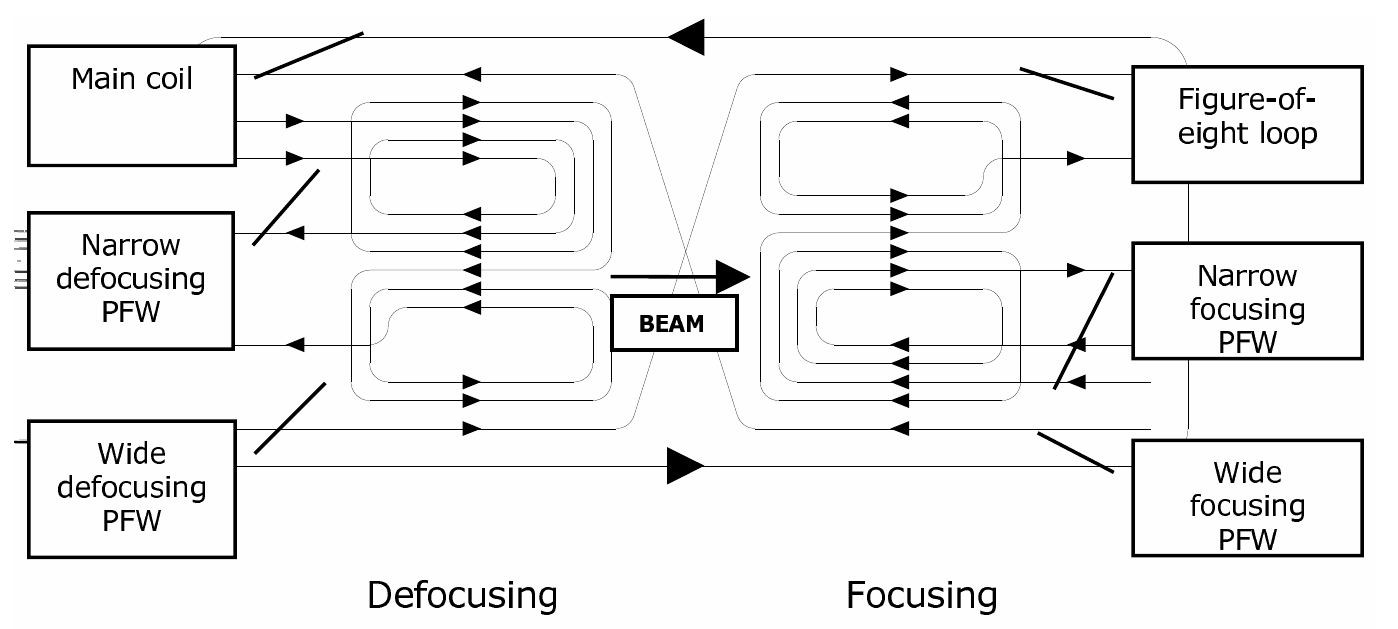}
\end{tabular}	
  \caption{\label{fig_a} Sketch of PS main magnet (left) and of the five circuits used to control the tunes and the chromaticities (right), from Ref.~\cite{marius}.}
\end{figure}

\begin{figure}[htb]
\centering
  \includegraphics[trim=0truemm 0truemm 0truemm 0truemm,width=0.4\linewidth,angle=0,clip=]{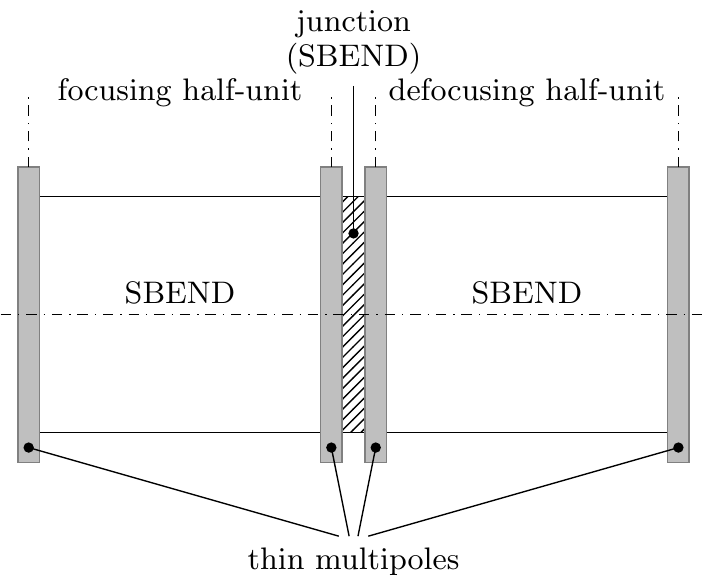}
  \caption{\label{fig_b}Schematic picture of the PS magnet model from~\cite{MTE-prog}. $K_1, K_2, K_3, K_4$ denote the quadrupole, sextupole, octupole and decapole thin lenses, respectively. The prefix D and F distinguish between the focusing and defocusing magnet families.}
\end{figure}

A sketch of the PS main magnet is shown in the left part of Fig.~\ref{fig_a}, while the schematic view of the five circuits to control tunes and chromaticities are shown in the right part.
In the PS numerical model, the dipole and quadrupole fields generated by the main combined function magnets are fixed by the beam momentum. The additional quadrupolar components induced by stray fields and PFWs are modelled by thin lenses placed at the half-unit ends. Sextupolar and higher-order fields generated by the PFWs and by the intermagnet gap are eventually represented by nonlinear thin lenses, also located at the half-unit ends. The thin lenses introduced are eventually grouped in two families, for focusing and defocusing half units, respectively, as sketched in Fig.~\ref{fig_b}. The PFWs affect global parameters such as tunes and chromaticities. Therefore, the impact on beam dynamics of the five circuits has been modelled by two {\sl effective} families of thin multipoles.
The effective PS lattice is computed by measuring the nonlinear chromaticity in both planes at the desired working point and energy. This beam-based technique has been implemented in the PS since 2002~\cite{PS-CHROM}. The transverse tunes are measured as functions of the momentum offset generated by a proper  radio-frequency perturbation. A polynomial fit of the measured curves is performed to extract numerical information on the different magnetic multipole orders. Then, the integrated strengths of the above thin lenses elements, $K_n=1/(B_0\rho_0)\, (\partial^nB_y/\partial x^n)$, where $B_0\rho_0$ is the magnetic rigidity, are computed to match the curves. This procedure is applied order by order, i.e. the quadrupole components are used to reproduce the constant term in the polynomial, the sextupole components the linear term, and so on, up to the decapolar components. The two families (defocusing and focusing) are used to match each quantity in both planes.

It is clear that the PS lattice is certainly not the easiest accelerator model to be used as example for the proposed scheme. It does not feature proper straight insertions, which would ensure the correction to zero of the dispersion and its derivative. Moreover, the fact that nonlinear magnetic fields are intrinsic to the additional coils wound in the main magnets implies that feed down effects cannot be avoided, which is also a complication for the proposed extraction or injection schemes. Nonetheless, this is not a serious obstacle to, at least, exemplify the novel concepts. 
\end{document}